\pdfoutput=1
\documentclass[format=manuscript, review=false, screen=true]{acmart}
\usepackage{booktabs} % For formal tables

\usepackage[ruled]{algorithm2e} % For algorithms
\usepackage{listings}
\usepackage[utf8]{inputenc}

\acmJournal{TRETS}
\setcopyright{acmlicensed}
\acmDOI{}

\acmYear{2019} 
\acmVolume{12} 
\acmNumber{4} 
\acmArticle{?} 
\acmMonth{9} 
\acmPrice{15.00}

\begin{document}
% Title portion
\title{GraVF-M: Graph Processing System Generation for Multi-FPGA Platforms}
%  \titlenote{This is a titlenote}
%  \subtitle{This is a subtitle}
%  \subtitlenote{Subtitle note}

\author{Nina Engelhardt}
\email{nengel@eee.hku.hk}
% \author{C.-H. Dominic Hung}
\author{Hayden K.-H. So}
\email{hso@eee.hku.hk}
\affiliation{%
  \department{Room 601 Chow Yei Ching Building, Department of Electrical and Electronic Engineering}
  \institution{University of Hong Kong}
  \streetaddress{Pokfulam Road}
  \city{Hong Kong}
}

\renewcommand\shortauthors{Engelhardt, So}

\begin{abstract}
Due to the irregular nature of connections in most graph datasets, partitioning graph analysis algorithms across multiple computational nodes that do not share a common memory inevitably leads to large amounts of interconnect traffic. Previous research has shown that FPGAs can outcompete software-based graph processing in shared memory contexts, but it remains an open question if this advantage can be maintained in distributed systems.

In this work, we present GraVF-M, a framework designed to ease the implementation of FPGA-based graph processing accelerators for multi-FPGA platforms with distributed memory. Based on a lightweight description of the algorithm kernel, the framework automatically generates optimized RTL code for the whole multi-FPGA design.
We exploit an aspect of the programming model to present a familiar message-passing paradigm to the user, while under the hood implementing a more efficient architecture that can reduce the necessary inter-FPGA network traffic by a factor equal to the average degree of the input graph.
A performance model based on a theoretical analysis of the factors influencing performance serves to evaluate the efficiency of our implementation. With a throughput of up to 5.8 GTEPS (billions of traversed edges per second) on a 4-FPGA system, the designs generated by GraVF-M compare favorably to state-of-the-art frameworks from the literature and reach 94\% of the projected performance limit of the system.
\end{abstract}

 \begin{CCSXML}
<ccs2012>
<concept>
<concept_id>10010520.10010521.10010542.10010543</concept_id>
<concept_desc>Computer systems organization~Reconfigurable computing</concept_desc>
<concept_significance>500</concept_significance>
</concept>
<concept>
<concept_id>10010583.10010600.10010628.10010629</concept_id>
<concept_desc>Hardware~Hardware accelerators</concept_desc>
<concept_significance>500</concept_significance>
</concept>
<concept>
<concept_id>10010583.10010600.10010628.10011716</concept_id>
<concept_desc>Hardware~Reconfigurable logic applications</concept_desc>
<concept_significance>500</concept_significance>
</concept>
</ccs2012>
\end{CCSXML}

\ccsdesc[500]{Computer systems organization~Reconfigurable computing}
\ccsdesc[500]{Hardware~Hardware accelerators}
\ccsdesc[500]{Hardware~Reconfigurable logic applications}


\keywords{Vertex Centric, Graph Processing, Multi-FPGA Architecture, Performance Modelling, FPGA, GraVF-M}

\maketitle
\thispagestyle{empty}
% Page limit 28 pages including references

\section{Introduction}

Graph data structures, which represent connections or relations between entities, are a natural fit for many data analysis applications. Popular uses range from determining social network influence drivers to finding the shortest route across a map of roads.
However, graph structures, and the algorithms that work on them, have unique characteristics that present challenges:

\begin{description}
    \item[Low computational density.] Most graph algorithms only execute a few operations per vertex or edge. Generally, the performance of a single processing node strongly depends on the speed at which graph data can be accessed.
    \item[Irregular structure.]Exacerbating the previous point, traversing a graph leads to random data access patterns with very low locality, which significantly degrades DRAM performance.
    \item[High degree of interconnection.] When partitioning a graph among multiple nodes, many edges will cross partitions, leading to high communication between nodes. While a good partitioning algorithm can mitigate this aspect somewhat, the communication bandwidth required can quickly rival the memory bandwidth and limit scaling of distributed graph algorithms.
\end{description}

FPGA-based accelerators can address these factors. Their on-chip BRAMs offer high-bandwidth low-latency random access performance, and their programmable fabric can implement custom architectures that efficiently use the available resources. Past research includes several specialized FPGA implementations of individual algorithms demonstrating the efficiencies that can be gained by approaches such as using BRAM to cache most frequently accessed vertex data \cite{zhang2018degreeawarehmc} or exploiting high-bandwidth off-chip memory interfaces \cite{attia2015apsp}. 

However, FPGAs are notoriously challenging to program. In addition to the domain knowledge required to be able to formulate the problem as an effective graph application, an FPGA implementation demands architecture and systems proficiency. Implementing a dedicated FPGA solution for a single, static application is therefore not practical in many contexts. Graph processing frameworks can encapsulate all the architecture and systems optimizations, putting FPGA systems within the reach of a domain expert with only basic hardware programming knowledge.

Research has explored approaches for more generic frameworks to accelerate a wider selection of graph algorithms. These leverage techniques from transactional memory \cite{ma2017transactional}, over custom soft processors \cite{kapre2015sparsegraph}, to CGRA overlays \cite{zhou2017tunao}. 
However, they mostly focus on single-FPGA accelerators, or on their integration with the software running on the host CPUs. Only a few works consider distributing a single computation across multiple FPGA boards. The popularity of the Convey HC-1/HC-2 (comprising four FPGAs) at the beginning of the decade prompted a number of works implementing some specific algorithm on this platform \cite{convey2011graph500,betkaoui2011framework}; but these primarily used the high-bandwidth shared memory for communication. There has been some investigation into the issues for true distributed multi-FPGA frameworks \cite{delorimier2006graphstep, dai2017multifpga}, but these have not yet been put to the test in an actual networked implementation.

It is an open question whether there is any performance advantage to be gained by scaling graph algorithms across multiple FPGAs. Due to the highly interconnected nature of graphs, distributed graph processing requires large communication volume between partitions. 
In our previous work, we have been investigating automatic generation of graph processing FPGA designs, and how we can maximize performance while minimizing the required user input \cite{engelhardt2016gravf, engelhardt2017HEART, engelhardt2018perfmodel}. A performance analysis revealed that this work is efficient on single-FPGA systems but suffers from communication inefficency when using lower-bandwidth off-chip networks.

This work introduces GraVF-M, a variant of GraVF optimized for multi-FPGA platforms. It improves on all aspects of the previous work by offering the following new contributions:
\begin{itemize}
    \item A reorganized architecture that significantly reduces communication over the inter-FPGA network. As network bandwith is the critical factor limiting multi-FPGA performance on most platforms, this directly corresponds to an equivalent speedup in total system performance.
    \item A three-stage programming model, which enables this optimization, as well as offering flexibility to the user and enabling simplified synchronization between supersteps.
    \item Better load balancing among PEs and FPGAs through low-overhead partitioning heuristics.
    \item An updated performance model and analysis whose predictions closely match the observed results.
\end{itemize}

Organization of this paper: 
Section~\ref{sec:related} situates our work in the context of previous efforts on FPGA-based graph accelerators. 
Section~\ref{sec:progmodel} explains the programming model for the user kernels and section~\ref{sec:arch} details our framework's implementation.
Section~\ref{sec:model} describes how we model system performance. 
Section~\ref{sec:results} evaluates a concrete implementation of our system on a 4-FPGA platform and compares it to the predictions of our model and to other works.
Finally, the conclusion is in section~\ref{sec:conclusion}.

\section{Related Work}
\label{sec:related}

Previous research on graph processing on FPGAs can be broadly categorized into two groups: efforts exploring an individual algorithm or platform, and efforts to build frameworks to more generally facilitate the implemenatation of graph algorithms.

Implementations of specific algorithms abound \cite{wang2010messagepassing, betkaoui2012BFS, convey2012graph500, attia2014cygraph, lei2015torus, umuroglu2015zyncbfs, dai2016fpgp, zhang2017bfshmc, khoram2018bfshmc, zhang2018degreeawarehmc}, but it is rare that an application is at the same time so business-essential and yet its requirements so slowly changing as to merit the several months of effort to build a dedicated hardware implementation.

Ours is not the first framework proposed to ease the implementation of graph processing on FPGAs. We have identified the following previous work introducing tools for FPGA-based graph processing not limited to a single application:

\subsection{Pure FPGA processing}

TuNao \cite{zhou2017tunao} is an ECGRA-based accelerator whose main optimization consists in storing the highest-degree vertices on-chip while leaving less highly connected vertices in off-chip memory. The accelerator targets a 3-phase Gather-Apply-Scatter execution model with a single data stream, parallelism is exploited by pipelining. The user kernel for each phase is implemented by a separate ECGRA module.

\cite{ma2017transactional} propose a transactional memory-based approach. Their model is GPU-like in that each processing element hosts a large number of threads that are frequently stalled due to high-latency shared memory accesses, however as the SIMD model is not effective for graph applications these processing elements will only process a single work item (e.g. a vertex kernel) at a time. Communication between different vertices happens via shared transactional memory, on the basis that most accesses will not conflict.

\cite{zhouprasanna2016edge, zhouprasanna2018edgecentric} argue that edge-centric graph processing is superior to the vertex-centric paradigm; they build an edge-centric accelerator. The first streams directly from off-chip memory, the second partitions the graph into multiple sections which can be processed in parallel by multiple PEs. Both designs rely on on strict sorting of the edge list for optimizations of the bandwidth needed to deliver messages.

\subsection{FPGA processing with access to host memory}

GraphGen \cite{nurvitadhi2014graphgen} is a graph accelerator built on the CoRAM memory interface, which gives the FPGA access to the host's main memory. The accelerator is a single processing element, that uses pipelining and SIMD processing to extract parallelism from the application. This is sufficient because of the low bandwidth connection to the main memory, but for systems where the graph data can be accessed more efficiently, the irregularity of the graph interferes with SIMD approaches.

GraphSoC \cite{kapre2015sparsegraph} is a custom soft processor-based approach where a vertex-centric algorithm kernel is expressed in 4 phases, and each kernel phase is implemented as a custom instruction. Thus the processors always execute the same program, but the effect of the instructions is changed. The processors exchange messages through an on-chip network, which also serves to access graph data from the host's main memory.

\subsection{CPU-FPGA hybrid processing}

GraphOps \cite{oguntebi2016graphops} proposes a library of dataflow actors for graph processing. They also propose a novel data structure that directly stores the vertex data of neighbors rather than an edgelist (which is a list of references to the neighbors). However, this requires the host to re-create this data structure for each iteration, and also increases the size of the graph storage.

\cite{zhouprasanna2017vertexedge} proposes a hybrid accelerator that is able to switch between vertex-centric and edge-centric execution depending on the current workload. The graph is partitioned in intervals and the CPU and FPGA both process different intervals in parallel (the FPGA with greater throughput than the CPU).

ExtraV \cite{Lee2017extrav} considers the case where the edge data is too large to fit into the host's main memory and has to be accessed from disk, but the vertex data is able to reside in DRAM. It uses the FPGA as an enhanced disk controller with near-data processing capabilities. The FPGA can read the edge data from disk in a compressed format, decompress and filter it on the fly and send the resulting list of neighbors to the host. The host then performs all algorithm-specific computation.

\subsection{Multi-FPGA capable frameworks.}

GraphStep \cite{delorimier2006graphstep, delorimier2011graphstep} was an early proponent of using the high internal memory bandwidth delivered by Block RAM to accelerate graph processing on FPGAs. They also proposed connecting multiple FPGAs together to increase the size of graph that can be processed. However, FPGA devices at the time offered even lower amounts of embedded memory than today, so their approach of storing the entire graph structure in BRAM resulted in (simulated) systems with hundreds of FPGAs for even modestly large graph datasets.

ForeGraph \cite{dai2017multifpga} is a multi-FPGA system targeting the Microsoft Catapult platform. It also consists of a number of PEs that communicate via a network.

Both of these frameworks' multi-FPGA performance is only evaluated in simulation. In contrast, our framework is fully implemented and performance numbers are actually observed.

\section{Programming model}
\label{sec:progmodel}

Based on a short user description, the GraVF-M framework generates a hardware description of a graph application for synthesis on platforms comprising one or more FPGAs as well as (optionally) off-chip memory.

The user input to the framework is a graph algorithm. The programming model used for this input has to fulfill double aims: to enhance productivity through ease of use, and to expose the algorithm features in a manner suited to hardware implementation.
We chose a synchronous vertex-centric programming model, and add some constraints related to data accesses to ensure a uniform interface for the resulting hardware kernel. Compared to our previous work \cite{engelhardt2016gravf}, the main difference is the addition of a gather stage, which aligns our programming model with the familiar gather-apply-scatter paradigm, and makes it easier for the programmer to ascertain that the mentioned constraints are respected.

In a synchronous vertex-centric programming model, the algorithm is formulated as a small function called a \emph{vertex kernel}. This user-provided kernel is run concurrently for each active vertex in the graph in what is called a \emph{superstep}. During a superstep, a vertex kernel only has access to limited data local to the vertex. Data is exchanged with neighboring vertices through messages. Supersteps are separated by global barrier synchronization, and any messages sent in superstep $n$ will only be received in superstep $n+1$.

This type of programming model is widely used as it is straightforward to understand, and any algorithm formulated in this manner is inherently embarrassingly parallel. However, it presents two main challenges due to the global barrier: The first is that imbalance in workload leads to time wasted waiting for stragglers to reach the barrier. This is mainly addressed through implementation choices that reduce imbalance and further mitigated by the use of a floating barrier. As these do not require modifications to the programming model, they will be discussed in the architecture section. The second challenge is that all messages generated in a superstep need to be stored until the next superstep. Memory resources are very limited on FPGA environments: typically, there will be less than 40MB on-chip BRAM and some 4-16GB off-chip DDR available.

To ensure correct and deadlock-free completion of algorithms, given the fixed-size buffers available, the quantity of message data generated in a superstep needs to be strictly bounded. We design the GraVF-M programming model so as to intrinsically enforce these limits, by asking the user to split their kernel into three functions (hardware modules) with fixed I/O interfaces.

\emph{Gather.} This function is called once for each message received by a vertex. It updates the vertex state based on the contents of the message.

\emph{Apply.} This function is called once for each vertex at the end of a superstep, after all messages have been received. Based on the final vertex state, it can issue an update to be communicated to the vertex's neighbors.

\emph{Scatter.} This function is called once for every outgoing neighbor. Based on the update issued by apply and, if defined, the edge data (weight), it finalizes the message to be sent to the neighbor.

Because the apply step is only called once for each vertex per superstep, the number of updates per superstep cannot exceed the number of vertices. Rather than attempt to store the much greater number of messages, we store these updates and only call the scatter function to produce the actual messages on demand when resources are available during execution of the next superstep. The messages are then generated, transported to their destination, and immediately consumed by the gather stage of the receiving vertex without risk of deadlock.

As an example, we show how the weakly connected components (WCC) algorithm is implemented in GraVF-M. This algorithm detects disjunct areas of a graph (sets of vertices such that all edges originating from a vertex in the set connect to a vertex also in the same set). These sets are also known as weakly connected components (to differentiate from strongly connected components, another name for cliques, where every vertex in the set is connected to every other vertex in the set). Each set agrees on a common label by propagating the lowest seen vertex ID to all neighbors.

The GraVF-M framework itself is written in Migen~\cite{migen}, a Python-based hardware description language that exports to Verilog for synthesis with conventional FPGA vendor tools such as Vivado from Xilinx. The native way to implement the algorithm kernels is also in Migen, but glue code is provided to insert any netlist that the FPGA vendor's tool accepts, thus the user is free to chose their favorite hardware description language.

In a first step, the user needs to define the algorithm-dependent data structures. These are defined in Migen/Python, as a list of field names and their bit widths. Predefined values related to system configuration, such as the size of a vertex identifier \lstinline"vertexidsize", can be used. The user can also define their own identifiers, whose values have to be provided in the configuration file at build. The following data structures need to be defined:

\begin{enumerate}
\item Vertex state data. For WCC, each vertex saves the smallest vertex ID found so far. The field \lstinline"active" indicates if a shorter path was found in this superstep that should be broadcast to neighbors.
\begin{lstlisting}[frame=none]
node_storage_layout = [
    ("label", "vertexidsize"),
    ("active", 1)
]
\end{lstlisting}

\item Edge data (optional). WCC does not consider edge weights, so this definition is omitted.

\item Update payload. When a smaller vertex ID is found, the information that the apply kernel needs to pass to the scatter kernel is the new label.
\begin{lstlisting}[frame=none]
update_layout = [
    ("label", "vertexidsize")
]
\end{lstlisting}

\item Message payload. The information sent by the scatter kernel to each neighbor is the new, smaller vertex ID encountered.
\begin{lstlisting}[frame=none]
message_layout = [
    ("label", "vertexidsize")
]
\end{lstlisting}

It frequently happens, as here, that the update payload and the message payload share the same format. In this case it is not necessary to write out the full definition again. These are simple Python variable definitions, so previously defined variables can be used:
\begin{lstlisting}[frame=none]
message_layout = update_layout
\end{lstlisting}
\end{enumerate}

In a second step, the three kernel modules Gather, Apply, and Scatter have to be defined, this time in the hardware description language of the user's choice. This example will use Verilog kernels, as more readers are likely to be familiar with this language than with Migen.

The data formats defined in the previous step are used to set the input and output signals of each module. All modules may be pipelined as deeply as the programmer wishes or have variable latency; they use handshake signals for flexible flow control.

\paragraph{Gather}
The gather kernel module receives as inputs a message from a neighbor that discovered a new label in the last superstep, and the destination vertex's state. The following inputs are defined:

\vspace{0.3em}
\begin{tabular}{l l}
\lstinline"level_in" & current superstep\\
\lstinline"nodeid_in" & active vertex's ID\\
\lstinline"sender_in" & neighbor that sent the received message\\
\lstinline"message_in" & received message (using \lstinline"message_layout")\\
\lstinline"state_in" & active vertex's current state data (using \lstinline"node_storage_layout")
\end{tabular}
\vspace{0.3em}

Flow control is handled by the associated handshake signals \lstinline"valid_in" and \lstinline"ready".
\vspace{0.3em}

The module should generate the following outputs:

\vspace{0.3em}
\begin{tabular}{l l}
\lstinline"nodeid_out" & active vertex's ID\\
\lstinline"state_out" & active vertex's new state data
\end{tabular}
\vspace{0.3em}

with the associated handshake signals \lstinline"state_valid" and \lstinline"state_ack".
\vspace{0.3em}

\begin{lstlisting}[caption={WCC gather implementation},label={src:gatherkernel}]
wire new_label;
assign new_label = state_in_label > message_in_label;

assign nodeid_out = nodeid_in;
assign state_out_label = new_label ? message_in_label : state_in_label;
assign state_out_active = new_label ? 1'b1 : state_in_active;
assign state_valid = valid_in;
assign ready = state_ack;
\end{lstlisting}

Listing \ref{src:gatherkernel} shows the implementation of the WCC gather kernel module. The amount of logic needed is very small, so it can be implemented using combinatorial logic only. A single comparison is needed: if the label from the message is smaller than the currently found smallest label, the label is updated in the vertex state and the vertex is marked active. Otherwise, the state is not modified.

% \begin{lstlisting}[caption={SSSP gather implementation},label={src:gatherkernel}]
% self.comb += [
%     If(self.state_in.dist > self.message_in.dist,
%         self.state_out.dist.eq(self.message_in.dist),
%         self.state_out.parent.eq(self.sender_in),
%         self.state_out.active.eq(1),
%     ).Else(
%         self.state_out.dist.eq(self.state_in.dist),
%         self.state_out.parent.eq(self.state_in.parent),
%         self.state_out.active.eq(self.state_in.active),
%     ),
%     self.nodeid_out.eq(self.nodeid_in),
%     self.state_valid.eq(self.valid_in),
%     self.ready.eq(self.state_ack)
% ]
% \end{lstlisting}

% \begin{lstlisting}[caption={SSSP gather implementation},label={src:gatherkernel}]
% wire new_path;
% assign new_path = state_in_dist > message_in_dist;

% assign nodeid_out = nodeid_in;
% assign state_out_dist = new_path ? message_in_dist : state_in_dist;
% assign state_out_parent = new_path ? sender_in : state_in_parent;
% assign state_out_active = new_path ? 1'b1 : state_in_active;
% assign state_valid = valid_in;
% assign ready = state_ack;
% \end{lstlisting}

\paragraph{Apply}
The apply kernel module processes each vertex's state at the end of the superstep, and initiates a broadcast of the new label if a smaller one was found this superstep. The following inputs are defined:

\vspace{0.3em}
\begin{tabular}{l l}
\lstinline"nodeid_in" & active vertex\\
\lstinline"state_in" & active vertex's current state data (using \lstinline"node_storage_layout")\\
\lstinline"round_in" & channel to use for messages sent in this superstep\\
\lstinline"barrier_in" & barrier marker (asserted after the last active vertex in the superstep)
\end{tabular}
\vspace{0.2em}

Flow control is handled by the associated handshake signals \lstinline"valid_in" and \lstinline"ready".

\vspace{0.3em}
This module should generate two sets of outputs:

\vspace{0.3em}
\begin{tabular}{l l}
\lstinline"nodeid_out" & active vertex\\
\lstinline"state_out" & active vertex's new state data (using \lstinline"node_storage_layout")\\
\lstinline"state_barrier" & barrier marker (indicates all state writebacks of a superstep are completed) \\
\lstinline"state_valid" & signals current output should be written back (no handshake)
\end{tabular}

There is no handshake for this output. To avoid the added complexity of handling two sources of backpressure, it is guaranteed that the state can be written back at any time.

\vspace{0.3em}
\begin{tabular}{l l}
\lstinline"update_out" & update to be broadcast (using \lstinline"update_layout")\\
\lstinline"update_sender" & vertex that generated the update\\
\lstinline"update_round" & channel to use for sending messages generated from the update\\
\lstinline"barrier_out" & barrier marker (separates updates of different supersteps)
\end{tabular}
\vspace{0.2em}

Flow control is handled by the associated handshake signals \lstinline"update_valid" and \lstinline"update_ack".
\vspace{0.3em}

Listing \ref{src:applykernel} shows the implementation of the WCC apply kernel module. Again, the logic is minimal: If the vertex has been marked active in the gather step, meaning that a smaller label has been found in this superstep, then an update is issued by asserting \lstinline"update_valid". All vertices also have their active bit reset to zero in preparation for the next superstep.

% \begin{lstlisting}[caption={SSSP apply implementation},label={src:applykernel}]
% self.comb += [
%     self.nodeid_out.eq(self.nodeid_in),
%     self.state_out.dist.eq(self.state_in.dist),
%     self.state_out.parent.eq(self.state_in.parent),
%     self.state_out.active.eq(0),
%     self.state_barrier.eq(self.barrier_in),
%     self.state_valid.eq(self.valid_in),
%     self.update_out.dist.eq(self.state_in.dist),
%     self.update_sender.eq(self.nodeid_in),
%     self.update_round.eq(self.round_in),
%     self.update_valid.eq(self.valid_in & self.state_in.active),
%     self.barrier_out.eq(self.barrier_in),
%     self.ready.eq(self.update_ack)
% ]
% \end{lstlisting}

% \begin{lstlisting}[caption={SSSP apply implementation},label={src:applykernel}]
% assign nodeid_out = nodeid_in;
% assign state_out_dist = state_in_dist;
% assign state_out_parent = state_in_parent;
% assign state_out_active = 1'b0;
% assign state_barrier = barrier_in;
% assign state_valid = valid_in;
% assign update_out_dist = state_in_dist;
% assign update_sender = nodeid_in;
% assign update_round = round_in;
% assign update_valid = valid_in & state_in_active;
% assign barrier_out = barrier_in;
% assign ready = update_ack;
% \end{lstlisting}

\begin{lstlisting}[caption={WCC apply implementation},label={src:applykernel}]
assign nodeid_out = nodeid_in;
assign state_out_label = state_in_label;
assign state_out_active = 1'b0;
assign state_barrier = barrier_in;
assign state_valid = valid_in;
assign update_out_label = state_in_label;
assign update_sender = nodeid_in;
assign update_round = round_in;
assign update_valid = valid_in & state_in_active;
assign barrier_out = barrier_in;
assign ready = update_ack;
\end{lstlisting}

\paragraph{Scatter}

The scatter kernel module processes the outgoing edges of a vertex that issued an update in the apply phase. For each edge, it propagates the label from the update.

The following inputs correspond directly to the update outputs from the apply kernel module (the same update is repeated multiple times for the different outgoing edges):

\vspace{0.3em}
\begin{tabular}{l l}
\lstinline"update_in" & update to be broadcast (using \lstinline"update_layout")\\
\lstinline"sender_in" & vertex that generated the update\\
\lstinline"round_in" & channel to use for sending messages generated from the update\\
\lstinline"barrier_in" & barrier marker (separates updates of different supersteps)
\end{tabular}
\vspace{0.3em}

Furthermore, the following inputs relate to each edge:

\vspace{0.3em}
\begin{tabular}{l l}
\lstinline"neighbor_in" & destination of the current edge\\
\lstinline"num_neighbors_in" & out-degree of the sending node
% \lstinline"edgedata_in" & data associated with the current edge (using \lstinline"edge_storage_layout")
\end{tabular}
\vspace{0.3em}

Input handshake is accomplished by the signals \lstinline"valid_in" and \lstinline"ready".

\vspace{0.5em}
The module outputs messages, using the following signals:

\vspace{0.3em}
\begin{tabular}{l l}
\lstinline"message_out" & message (using \lstinline"message_layout")\\
\lstinline"neighbor_out" & destination vertex\\
\lstinline"sender_out" & sending vertex\\
\lstinline"round_out" & channel to use for sending the message\\
\lstinline"barrier_out" & barrier marker (separates messages of different supersteps)
\end{tabular}
\vspace{0.3em}

Output handshake signals are \lstinline"valid_out" and \lstinline"message_ack".

Listing \ref{src:scatterkernel} shows the scatter kernel module for WCC. All values are forwarded as-is. While it is not necessary for these small amounts of logic, we insert a pipeline stage to demonstrate how a pipelined module would be implemented.

% \begin{lstlisting}[caption={SSSP scatter implementation},label={src:scatterkernel}]
% self.sync += If(self.message_ack,
%     self.message_out.dist.eq(self.update_in.dist + self.edgedata_in.dist),
%     self.neighbor_out.eq(self.neighbor_in),
%     self.sender_out.eq(self.sender_in),
%     self.round_out.eq(self.round_in),
%     self.valid_out.eq(self.valid_in),
%     self.barrier_out.eq(self.barrier_in)
% )
% self.comb += self.ready.eq(self.message_ack)
% \end{lstlisting}
% \begin{lstlisting}[caption={SSSP scatter implementation},label={src:scatterkernel}]
% assign ready = message_ack;

% always @(posedge sys_clk) begin
% 	if (message_ack) begin
% 		message_out_dist <= (update_in_dist + edgedata_in_dist);
% 		neighbor_out <= neighbor_in;
% 		sender_out <= sender_in;
% 		round_out <= round_in;
% 		valid_out <= valid_in;
% 		barrier_out <= barrier_in;
% 	end
% end
% \end{lstlisting}

\begin{lstlisting}[caption={WCC scatter implementation},label={src:scatterkernel}]
assign ready = message_ack;

always @(posedge sys_clk) begin
	if (message_ack) begin
		message_out_label <= update_in_label;
		neighbor_out <= neighbor_in;
		sender_out <= sender_in;
		round_out <= round_in;
		valid_out <= valid_in;
		barrier_out <= barrier_in;
	end
end
\end{lstlisting}

Cumulatively, the whole hardware description of the WCC algorithm is implemented using approximately 30 lines of code. This is typical of most kernels, as long as they only use integer arithmetic. An algorithm with multiple floating-point operations, such as e.g. PageRank, requires about 200 lines of code to implement the pipelining necessary for efficient processing.

\paragraph{Type of algorithms supported}
Not all graph algorithms can be implemented in the Gather-Apply-Scatter programming model chosen. An interesting way to view the restrictions is by considering the matrix representation of graphs: Instead of representing a graph as a collection of vertices and edges, it is also possible to represent a graph as an adjacency matrix of size $|V|\times |V|$, where the element $A_{ij} = 1$ if the vertices $i$ and $j$ are connected by an edge, and $A_{ij} = 0$ otherwise. The data accesses performed in this Gather-Apply-Scatter model are the same as for multiplying the vector of vertex states with the adjacency matrix, and saving the result as the new vertex state.

In terms of the equivalent matrix representation of graphs, the algorithms that can be implemented are of matrix-vector type: throughout all supersteps, the resulting vector of vertex data remains of constant size. Operations of type matrix-matrix, or in graph terms operations that build lists of multi-hop paths, are deliberately excluded since they have dynamically increasing storage requirements (the result of multiplying a sparse matrix with itself has more non-zero values than the original). Although there is a theoretical transformation that allows any vertex-centric algorithm to be turned into a 3-phase Gather-Apply-Scatter kernel, these transformations rely on non-bounded data structures (e.g., keeping all messages received in the Gather stage saved in the vertex data, and only using them in the Apply stage). These cannot be realistically applied on FPGA, which has much more limited and less flexible memory resources.

%We have developed a variant of GraVF that can implement a larger fraction of algorithms by relaxing the ``one update per vertex per superstep'' requirement (which requires merging the Gather and Apply steps into a single combined module). Without this restriction, the memory usage of the update queue that stores the (Gather-)Apply kernel outputs is no longer bounded, and it is incumbent on the user to avoid deadlock situations arising from exceeding the memory allocated to the queue. We have found this model much more difficult to use and the results obtained so far on algorithms that cannot fit the 3-phase Gather-Apply-Scatter model not compelling. The restrictions imposed by the 3-phase programming model can therefore be considered a screening process for the algorithms most likely to benefit from a GraVF-generated accelerator.
%\section{Automatic System Generation with GraVF}
\section{Architecture}
\label{sec:arch}

\begin{figure}[h]
\includegraphics[width = 0.5\linewidth]{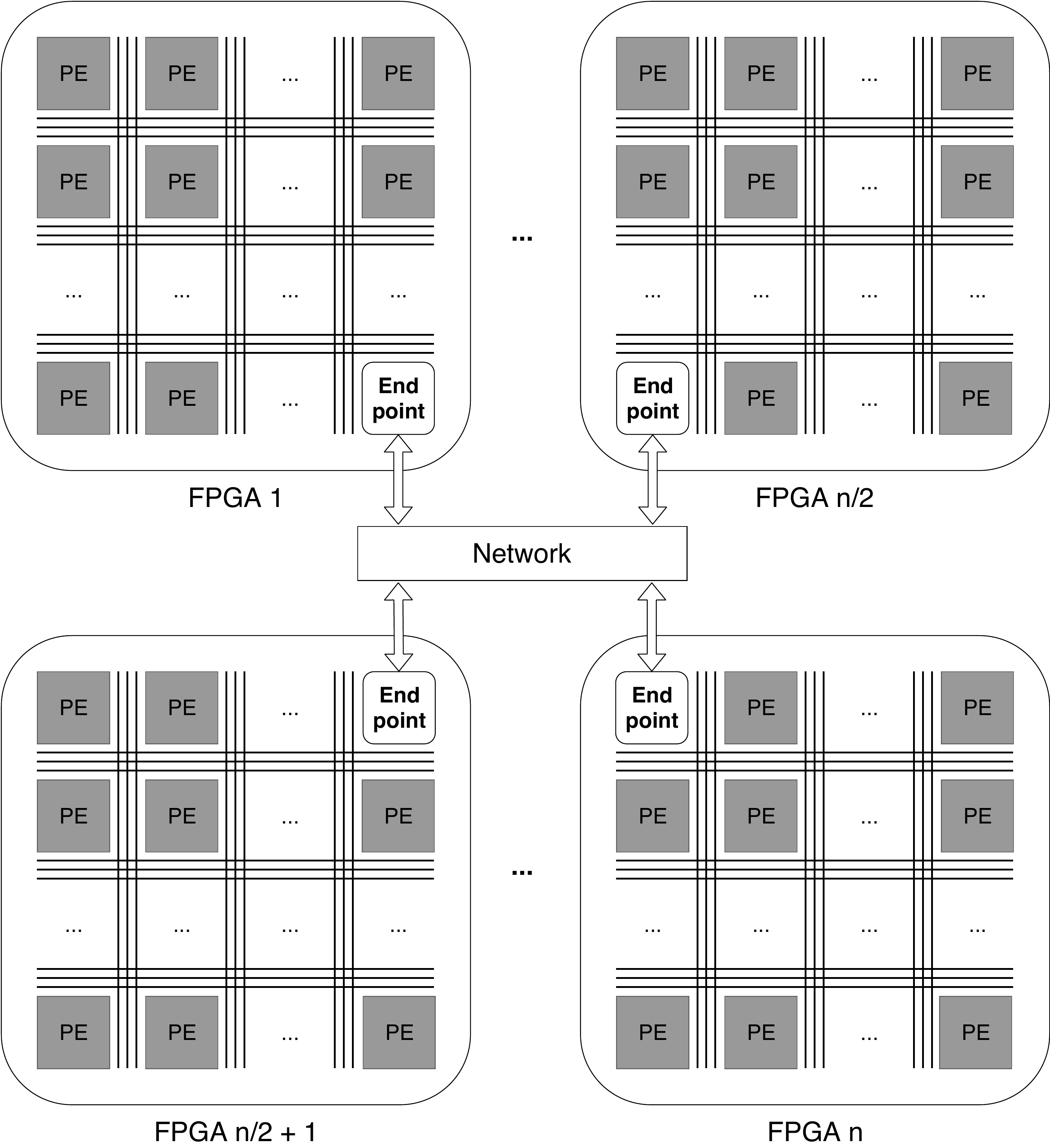}
\caption{System overview\label{fig:sys}}
\end{figure}

One of the advantages of using a vertex-centric model is that large amounts of parallelism are readily available. To make use of it, the GraVF-M framework generates as many processing elements (PEs) as can fit the available resources. The vertices of the graph are then partitioned among the PEs. The assigned PE of a vertex hosts the vertex's data and runs the vertex kernel for this vertex. Communication between vertices involves transfer of data between different PEs. An on-chip network, in our case realized as a crossbar, transports data between PEs on the same FPGA. On multi-FPGA systems, data destined for remote PEs is handed to a network endpoint that delivers it via off-chip network to the endpoint on the destination FPGA. Fig. \ref{fig:sys} shows a high-level overview of the complete system.

\subsection{Optimized message delivery for multi-FPGA systems}

When implementing a message-passing graph processing system, the amount of messages to be delivered is a concern in two ways: 

\begin{itemize}
\item Storage space. Because computation is structured into supersteps, all messages sent during a superstep have to be stored before they can be delivered to the next superstep. The available memory space puts an upper limit on the amount of messages that can be sent during a superstep, which generally correlates with the dataset size.
\item Communication bandwidth. Graphs are highly interconnected, usually in an unstructured manner, which means that partitioning the graph between multiple processing elements or across multiple FPGAs will lead to large amounts of messages being sent between partitions.
\end{itemize}

To address both issues, we introduce an optimization in GraVF-M which makes use of the observation that during the scatter phase, one update produces many messages.
We can save both memory space and bandwidth by waiting until the last possible moment to call the scatter function, which means producing messages on demand to be immediately consumed by the gather module.

From the user's perspective, the programming model is unchanged: they write the program as if messages were being exchanged by distinct supersteps. In the underlying hardware generated by the framework, however, the execution of the scatter stage is delayed until the following superstep, and it is located in the receiving PE.

The sending PE stores all updates produced by the apply stage during a superstep in a large output queue. Because the programming model is defined such that at most one update is allowed to be issued by each vertex in a superstep, the size of this queue is equal to the number of vertices which is generally at least an order of magnitude smaller than the number of messages, of which there can be as many as there are outgoing edges. Rather than creating individual messages and transporting them to their destination, the update is broadcast to all PEs, which then execute the scatter stage for the subset of edges that have local destinations.

\subsection{Processing Elements}

\begin{figure}[h]
\includegraphics[width = 0.7\linewidth]{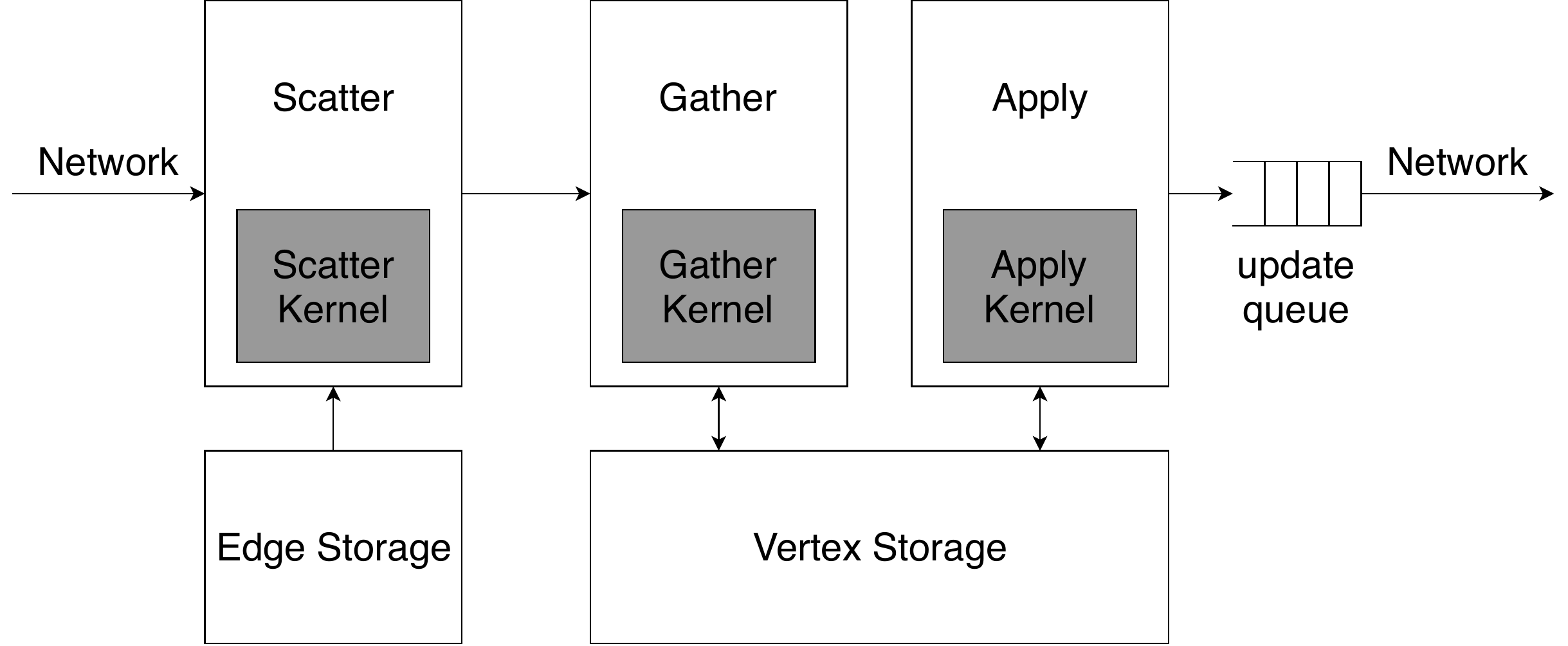}
\caption{Multi-FPGA optimized processing element overview\label{fig:pe}}
\end{figure}

The processing element is responsible for organizing the execution of the vertex kernels according to the described programming model. This means arranging the flow of messages, retrieving the associated vertex and edge data, and managing the update queue, which is the principal storage in the system.

Like the algorithm, the PEs are split into three modules for the three phases gather, apply and scatter. Fig. \ref{fig:pe} gives an overview of the PE's internal pipeline, with the user-provided kernel modules shaded.

\subsubsection{Scatter}
\begin{figure}[h]
\includegraphics[scale=0.6]{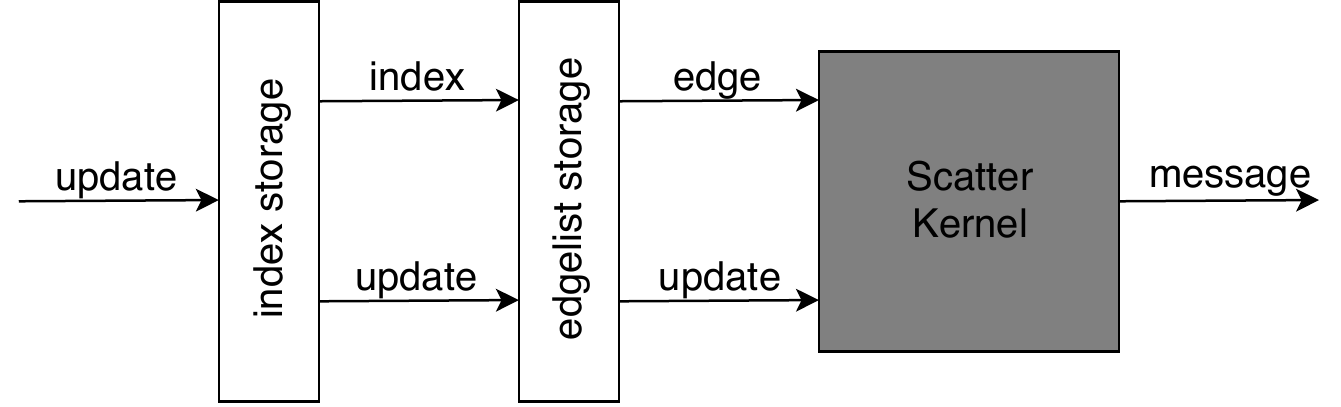}
\caption{Scatter module\label{fig:scatter}}
\end{figure}

Incoming updates from the previous superstep enter the scatter module (Fig.~\ref{fig:scatter}). For each update, the scatter module first looks up the starting address and length of the sender's local portion of the edge list, then iterates through the local neighbors and presents the edges one by one to the scatter kernel along with the update. Edgelist storage can be local (in BRAM) or in off-chip memory, such as HMC. The last update of a superstep is followed by a special barrier update, which is passed through the scatter kernel to mark the end of the messages.

\paragraph{Edge storage modifications}
\label{sec:scatter_modif}

\begin{figure}[h]
\includegraphics[width=0.3\linewidth]{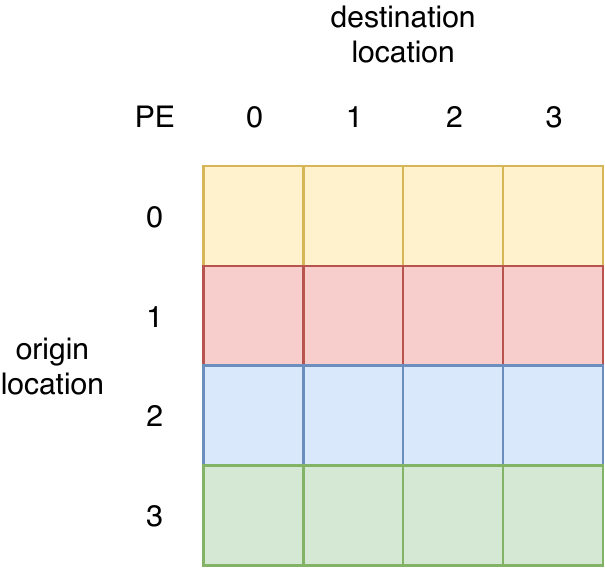}\hspace{4em}
\includegraphics[width=0.3\linewidth]{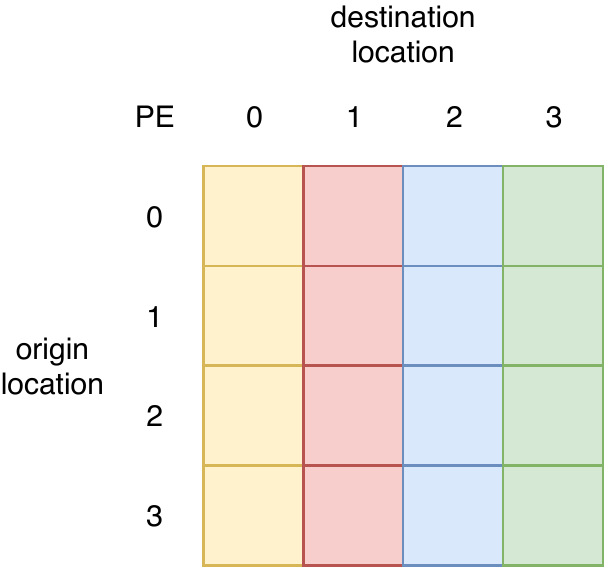}
\caption{Edge list partitioning in GraVF (left) and GraVF-M (right)}
\label{fig:edge_distrib}
\end{figure}

Delaying execution of the scatter stage necessitates changing how the edge data is stored.
% The scatter module itself is not modified, however the datasets in its edge storage are changed. 
Rather than storing the list of all neighbors for the subset of vertices assigned to this PE, it now stores for all vertices the subset of neighbors assigned to this PE. The overall volume of edges stored is the same, but they are moved to a different location. Fig. \ref{fig:edge_distrib} illustrates where edges are stored in the original (GraVF) and the new (GraVF-M) scheme for the case of 4 PEs, with each color representing the partition of edges stored in the same scatter module.

\subsubsection{Gather}

\begin{figure}[h]
\includegraphics[scale=0.5]{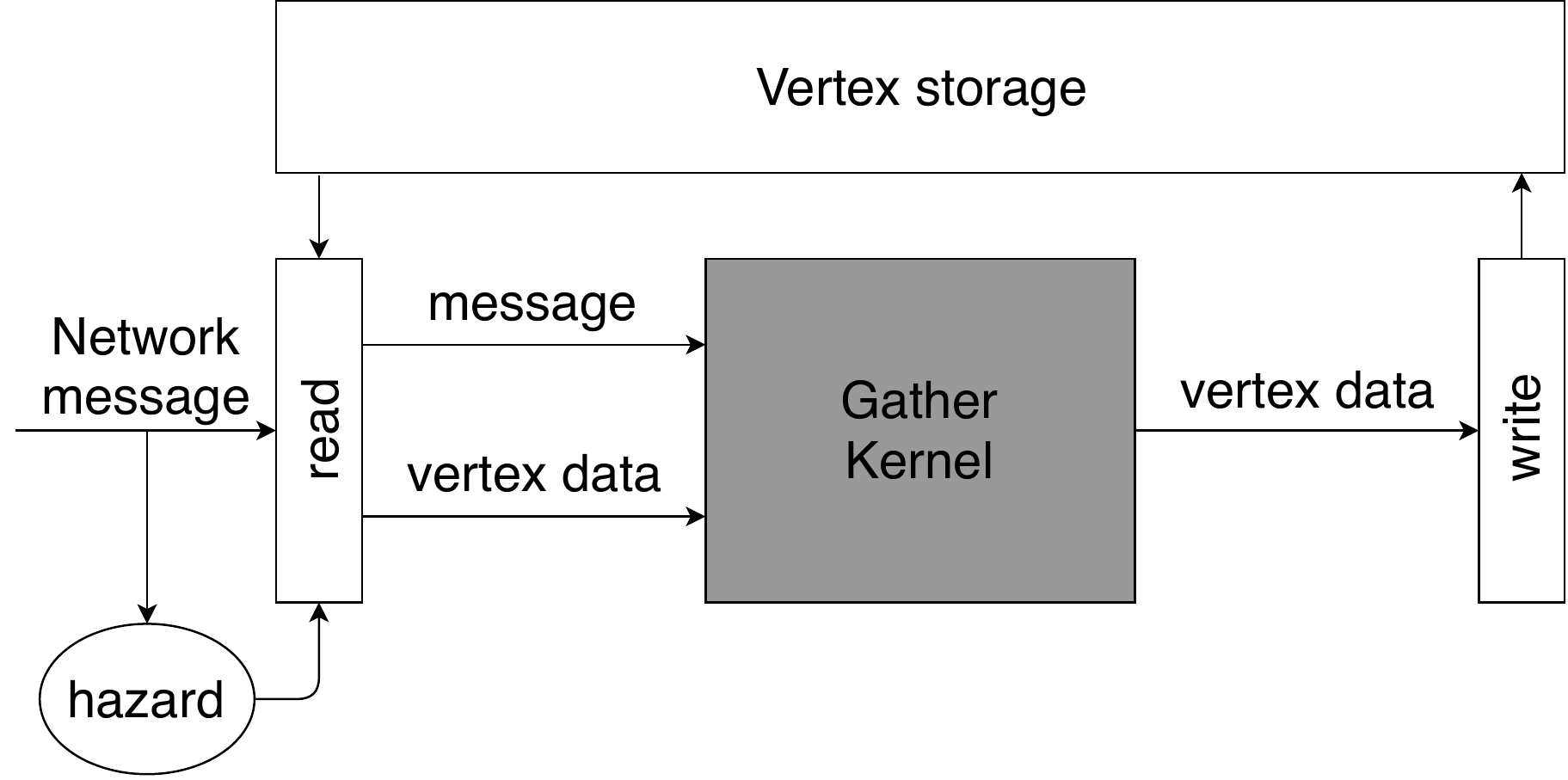}
\caption{Gather module\label{fig:gather}}
\end{figure}

Messages produced by the scatter module enter the gather module, whose internals are shown in Fig. \ref{fig:gather}. The module reads the destination vertex's data, and passes it together with the message to the user-defined gather kernel. A read-after-write hazard may occur if another message for the same vertex arrives in quick succession, and the destination vertex's data has not yet been written back. We use a hazard detection module to keep track of in-use vertex data and stall the read stage until the correct data has been written back by the gather kernel.

\subsubsection{Apply}

\begin{figure}[h]
\includegraphics[scale=0.5]{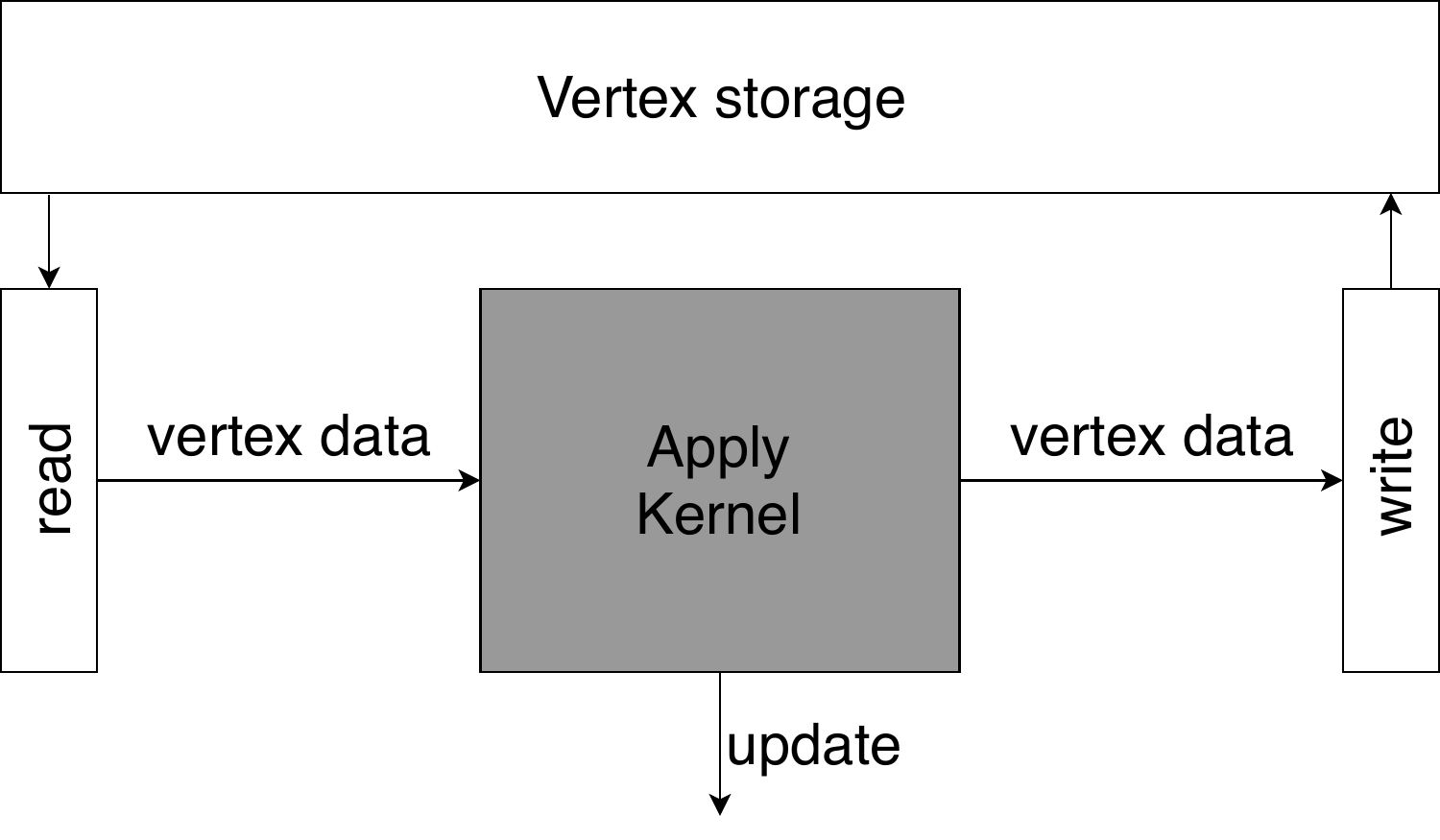}
\caption{Apply module\label{fig:apply}}
\end{figure}

When a barrier message enters the gather stage, this signals that all messages for the current superstep have been received. The ready signal from the gather module is deasserted to temporarily halt further transmission of messages from the next superstep that might potentially already be waiting.
The gather module is first flushed to ensure that all vertex data is written back, then access to the vertex storage is switched to the apply module (Fig. \ref{fig:apply}). The apply module iterates through all vertices assigned to the PE, passing them sequentially to the user-defined apply kernel. The updates generated by the apply kernel are entered into a large update queue. Once all modifications to vertex data are written back, a barrier update is written to the queue to separate updates of this superstep from those of the next, and control of the vertex data is switched back to the gather module, which once again begins accepting messages.

\subsection{Network and Synchronization}

The role of the network is not only to transport data between the PEs, but also to enforce separation between the supersteps. The programming model specifies a barrier between supersteps: all messages from one superstep should finish processing before the first message from the next superstep is processed. Given the deep pipelines necessary to make FPGA-based processing efficient and the distributed nature of the system with many processing elements advancing at different rates, enforcing such a strict temporal separation between supersteps would waste many cycles flushing the complete system at each superstep. Instead, we propose the use of a \emph{floating barrier} to preserve the invariant, from the user's perspective, that they will always receive all messages from one superstep before the first message from the next -- but different pipeline steps and different PEs may proceed to the next superstep at different times.

\paragraph{Floating Barrier}
To achieve flexible distributed barrier synchronization, barrier messages are exchanged between PEs. Initially, the barrier is injected into the apply modules to begin execution. This causes each apply module to call the apply kernel on the initial state of its assigned vertices, and then follow with the barrier. The barrier progresses through the pipeline behind the initial updates, terminating the data for this superstep. As soon as the apply module finishes writing back the modified vertex data of the last vertex, it switches operations back to the gather module, which is now ready to accept any messages for the next superstep. These are likely to be already waiting, as the earlier updates have had time to progress through the scatter module and cause messages to be generated. While the barrier remains in the update queue, the gather module is in effect processing a superstep ahead of the scatter module.

Throughout the PE pipeline, strict ordering is enforced, with the barrier following behind the data for the current superstep. When the barrier reaches the end of the apply module and exits the PE, the network sends a copy of the barrier to each PE individually, and includes the count of updates that was sent during this superstep to the destination PE in the barrier. This count allows the destination to confirm it has received all updates for this superstep, which is necessary to support networks which do not guarantee in-order delivery (e.g. packet-switched off-chip networks like Ethernet). If the barrier message were to be reordered and arrive before other messages, the destination network endpoints would wait for the remaining messages.

On the receiving side, each PE's network endpoint passes on updates to the gather module, and waits to receive a barrier from each PE in the system (including itself). Upon reception of all barriers and comparison of received and expected update counts, the network endpoint of the PE is able to independently determine that it has received all updates for this superstep. At this point the barrier is passed on to the PE, which can progress to the next superstep.

Detecting algorithm termination is also achieved through the barrier messages. The message includes a bit indicating whether the sending PE has sent any updates during the last superstep. If none of the PEs have generated any updates, the algorithm is deemed inactive and the receiving network endpoint does not pass the barrier on but raises a termination signal instead. This way all PEs are able to detect termination independently, in true distributed fashion.

\paragraph{Network properties}

For the floating barrier algorithm to work properly and not cause deadlock, the network needs to fulfill two conditions: it needs to guarantee delivery (i.e. not lose any updates), and it needs to guarantee that backpressure on updates of one superstep will not impede delivery of updates from any previous superstep.

The first condition is straightforward: if updates are lost in transit, the algorithm execution is no longer correct. Deadlock would also occur because the receiving PE would wait indefinitely for the updates to arrive. 

The second condition also serves to avoid deadlock. Because the floating barrier allows PEs to progress to the next superstep at different times, updates from different supersteps may coexist. Buffering capacity in the network is very limited. Therefore, if some PE advances to the next superstep and starts sending many messages before any of the other PEs are ready to accept them, these messages would fill all available buffers in the network. If this impedes delivery of any messages from the previous superstep that the other PEs are still waiting for, they will never be able to progress to the next superstep. However, the messages from the new superstep clogging the buffers cannot be consumed until after the other PEs switch to the next superstep. To ensure this deadlock situation cannot occur, the network must not allow messages from a later superstep to interfere with delivery of earlier messages.

The easiest way to implement this would be to only admit messages of one superstep into the network at a time, and prohibit sending messages of the next superstep until all messages of the previous superstep are delivered. However this reduces the benefit of the floating barrier, as it means that a true temporal barrier is introduced and variable workload imbalances between PEs can no longer be averaged out across supersteps. For better performance, the network should offer separate (physical or virtual) channels for messages of different supersteps.

We implement internal communication within the FPGA through a crossbar network. The receiving side indicates which channel it is currently accepting messages from, and the arbiter selects a sender among PEs wishing to send on this channel.

For inter-FPGA communication, we sequentialize the virtual channels. The endpoint for outgoing messages in the crossbar only accepts messages from one superstep at a time, and switches to the next channel when all PEs within the FPGA have completed the superstep.

\paragraph{Filtering delivery of updates to FPGAs with neighbors}
\label{sec:filter}

In comparison to GraVF, the main change to the network is that instead of unicast messages, it now transports broadcast updates that have to be delivered to all PEs in the system. The local network is modified in such a manner that updates sent from a local PE are broadcast to all local PEs and to each remote FPGA, whereas updates coming from remote FPGAs are broadcast only to local PEs. Updates, just like messsages, use channels to separate supersteps and are counted to ensure correct synchronization. Barriers work the same as previously to detect superstep switchover and algorithm termination.

Broadcasting an update to all PEs is only advantageous if an update, on average, produces more messages than there are copies made of the update. If most vertices have only one neighbor, sending a copy of the update to each other FPGA in the system will use more inter-FPGA bandwidth than a single message would. To ensure that this optimization will always result in a net benefit, we add a filtering unit that looks up whether the sender of an update has neighbors on a remote FPGA, and does not send it to FPGAs where there are no potential recipients. This is currently realized through a bitmap of size $|V|\times n_\text{FPGA}$ (listing for each vertex which FPGAs host neighbors) that is generated at the same time as the graph is partitioned across FPGAs. For larger systems, the storage required for the filter could be reduced through use of compression or heuristic methods such as bloom filters.

\subsection{Partitioning}

How the vertices are distributed among the PEs affects the workload balance. Better partitioning improves processing speed, however the pre-processing overhead can quickly overwhelm these gains. The least-effort way to distribute vertices between the PEs is by round-robin, which results in a balanced number of vertices per PE. However, as the workload consists primarily of traversing edges, the cumulative number of outgoing edges from the vertices assigned to a PE would be a better proxy for workload balance.

Balancing the number of edges per PE is an instance of the multi-way number partitioning problem and thus theoretically NP-complete, however this problem is known to be solved well by even the simplest of heuristics if the number of partitions is much smaller than the number of items to be partitioned~\cite{korf2009multiway}. In fact, we find that we achieve near-perfect partitioning by assigning vertices to the partition with the currently lowest edge count even if we omit the preliminary step of sorting the vertices by degree. This is presumably helped by the fact that the vertices are already somewhat sorted due to the way the graph is constructed when reading in the edge list: vertices are added as they are encountered, and vertices with many edges appear many times and thus have a higher likelihood of being encountered early.

When distributing vertices within a single FPGA, there are no locality benefits: the communication costs are identical between any two PEs, as even a message sent to another vertex on the same PE has to pass through the network so as to respect the superstep synchronization. However, once we introduce multiple FPGAs, there is a significant difference in available bandwidth for communicating with vertices located on remote FPGAs compared to local vertices.

We add the option to partition the graph between the FPGAs using METIS~\cite{karypis1998metis}. While the runtime of this high-quality partitioning method is several orders of magnitude larger than that of our benchmarks, it might still be worthwhile in specific circumstances (e.g. when the same dataset is reused many times), and it allows us to explore some of the tradeoffs in partitioning.

As before, we use the outdegree of vertices as weight for balancing. METIS allows us to tune a balance factor \lstinline"ufactor", which constrains the allowed imbalance of the resulting partition to be no greater than \lstinline"ufactor"/1000. The default value of \lstinline"ufactor" is 1 (0.1\% of imbalance allowed), and we observe that for RMAT graphs, the quality of the partition only improves for values greater than 500. As the performance degradation from imbalance is much greater than the gain from reducing inter-FPGA communication, we leave the value of \lstinline"ufactor" at 1.

\paragraph{Dataset size limitations}

So far, data is only partitioned in space, among the different PEs. We have not implemented partitioning in time, where different parts of the graph are loaded for processing one after the other by the same PE. The size of the datasets that GraVF-M can handle is consequently limited by the need to fit the entire graph in the resources provided by the FPGAs available. %Vertex data and the index structure of the edge lists is stored in Block RAM, of which an FPGA typically only offers a few tens of megabytes. Although internal memory resources are increasing (e.g. with availability of UltraRAM in newer Xilinx devices), so are graph sizes.

However, GraVF would be a good fit for grid-based partitioning, as is used e.g. in the software framework GridGraph~\cite{zhu2015grid} or in the other framework using multiple FPGAs, ForeGraph~\cite{dai2017multifpga}. In this approach, the edge lists are partitioned both by origin and by destination vertex, so that the scatter stage can generate messages separately for each block of vertices. This allows the gather stage to load the relevant vertices and process all messages destined for this block, and then proceed to load the next block.

\section{Theoretical Performance Analysis}
\label{sec:model}

We build a throughput-oriented performance model of the system based on an analysis of the various limiting factors. This model serves a dual purpose, both to project expected performance of a system before proceeding to acquire it and implement the algorithm, and for use by the framework during generation to adjust the system parameters to their optimal values.

\subsection{Objective}

The goal of this performance model is to maximise the overall system throughput ($T_\text{sys}$) by determining the values of the following two variables: the number of FPGAs ($n_\text{FPGA}$) and the number of PEs per FPGA ($n_\text{PE/FPGA}$). System throughput is measured in traversed edges per second (TEPS), a common measure of performance for graph processing.

% Traversing one edge requires the PE to make the following data transfers: it receives one message, loads one edge, and sends one message. As the number of PEs in an FPGA increases, either the step of loading the edges from memory or the sending and receiving of messages over the network becomes the rate-limiting factor.

\subsection{Parameters}

\begin{table}[h]
\begin{tabular}{ll}
\multicolumn{2}{l}{Variables} \\
\hline
$n_\text{FPGA}$ & number of FPGAs to use\\
$n_\text{PE/FPGA}$ & number of PEs per FPGA\\
\\
\multicolumn{2}{l}{Dependent Variables} \\
\hline
$T_\text{FPGA}$ & throughput per FPGA (edges/s)\\
$T_\text{sys}$ & total system throughput (edges/s)\\
\\
\multicolumn{2}{l}{System Parameters}\\
\hline
$CPE_\text{PE}$ & cycles per edge (edges$^{-1}$)\\
$f_\text{clk}$ & operating frequency (Hz)\\
$BW_\text{if}$ & network interface bandwidth (bits/s)\\
$BW_\text{network}$ & total network bandwidth (bits/s)\\
$BW_\text{mem}$ & memory interface bandwidth (bits/s)\\
$M_\text{board}$ & memory capacity per FPGA (bits)\\
\\
\multicolumn{2}{l}{Algorithm-dependent Parameters}\\
\hline
$m_\text{vertex}$ & data storage per vertex (bits)\\
$m_\text{message}$ & message size (bits/edge)\\
$m_\text{update}$ & update size (bits)\\
$m_\text{edge}$ & edge size (bits)\\
$p_\text{msg/TE}$ & messages per traversed edge (edges$^{-1}$)\\
\\
\multicolumn{2}{l}{Dataset-dependent Parameters}\\
\hline
$|V|$ & number of vertices in the input graph\\
$|E|$ & number of edges in the input graph\\
\end{tabular}
\vspace{1em}
\caption{Definitions}
\end{table}

The domains for the two variables $n_\text{FPGA}$ and $n_\text{PE/FPGA}$ are determined by the platform configuration. When including support for a given FPGA board in the framework, we specify the available resources as part of the board support package. The domain for the number of PEs per FPGA ($n_\text{PE/FPGA}$) is fixed based on the size of the reconfigurable fabric. The computational demands of graph algorithms are generally low, so that the area demands for the user-provided application kernels can be estimated generously. The upper limit is straightforwardly determined by reconfigurable logic element usage per PE. For larger FPGAs, a lower limit greater than one PE may also be imposed to avoid routing problems: as we strive to use all available BRAM to store vertex data, it becomes very hard for the synthesis tools to route connections from all corners of the FPGA chip to a single PE.

The minimum number of FPGA boards is determined by the problem size: each board can host a number of vertices limited by the available memory and the size of the vertex data specified by the algorithm. The number of FPGAs in the system has to be big enough to host the whole dataset:

\[n_\text{FPGA} \geq \frac{|V| \times m_\text{vertex}}{M_\text{board}}\]

Given these parameters, the system evaluates which of the following factors limits the overall performance:

\subsection{Processing element throughput constraint.} Each processing element can traverse a limited number of edges per second. Analogous to a processor's CPI (cycles per instruction), we define the processing element's CPE (cycles per edge) as the average number of cycles it takes to traverse an edge. It is determined by a combination of architectural features, input graph features and message patterns: The PE pipeline is for the most part designed to handle a new edge each cycle, but the gather module can stall due to read-after-write hazards if two messages for the same vertex are received in quick succession, and the scatter module needs a cycle to reset at the end of each edge list. The user kernel may also inadvertently lower CPE if an inefficient implementation is chosen, although this is not the case for our kernel implementations, which all support a CPE of 1. Experimentally, we determine the CPE of our PEs to vary between 1.05 and 1.4.

The cumulative throughput limit $L_{PE}$ of all PEs in the system is expressed in the following formula:
\begin{equation}
\label{eq:complim}
\boxed{
T_\text{sys} \leq L_{PE} = n_\text{FPGA} \times n_\text{PE/FPGA} \times \frac{f_\text{clk}}{\mathit{CPE}_\text{PE}}
}
\end{equation}
For the maximum number of PEs that fits in an FPGA, we obtain the computational limit of the FPGA, $L_{PE_{max}}$.

\subsection{Memory bandwidth constraint.}
When using off-chip memory to store the adjacency lists, the FPGA throughput $T_\text{FPGA}$ may also be limited by the memory bandwidth. Traversing one edge requires loading one edge of size $m_\text{edge}$ from memory:
    
\[T_\text{FPGA} \times m_\text{edge} \leq BW_\text{mem}\]
From which, by multiplying with the number of FPGAs in the system, we derive the memory interface limit $L_{mem}$:
\begin{equation}
\label{eq:memlim}
\boxed{
T_\text{sys} \leq L_{mem} = n_\text{FPGA} \times \frac{BW_\text{mem}}{m_\text{edge}}
}
\end{equation}

The above equation does not account for memory access granularity. While the HMC platform used in our experiments allows accessing memory in comparatively small words of 128 bit, leading to very little wasted bandwidth, this can become important e.g. when using AXI-based memory interfaces, which read data in words of 512 bit at a time. The effect is also more pronounced for GraVF-M than for GraVF, as the edgelists are split into $n_\text{PE}$ arrays, as discussed in section~\ref{sec:scatter_modif} and shown in Fig. \ref{fig:edge_distrib}.

We can estimate the worst case wasted bandwidth if we assume that there is no correlation between the length of the edgelist and the probability of it being accessed (i.e. vertices are equally likely to be active irrespective of their degree) as follows.

There are $|V|\times n_\text{PE}$ edgelists, containing $|E|$ edges in total. Each edgelist is stored as a sequential array of edges and accessed as a whole. The start of the edgelist is aligned to the memory access word boundaries. In the worst case, the last memory access for every edgelist contains only a single edge. Thus, to read all $|E|$ edges in the graph, on top of the useful $|E| \times m_\text{edge}$ bits of edgelist, an additional $|V| \times n_\text{PE} \times (m_\text{memword}-m_\text{edge})$ bits are read from memory (where $m_\text{memword}$ denotes the number of bits read in one memory access). In per-traversed-edge terms, this corresponds to:

\[T_\text{FPGA} \times \left(m_\text{edge} + \frac{|V|}{|E|} \times n_\text{PE} \times (m_\text{memword}-m_\text{edge})\right) \leq BW_\text{mem}\]

From which we can derive the refined limit accounting for access granularity:

\[
T_\text{sys} \leq \frac{n_\text{FPGA} \times BW_\text{mem}}{\left(m_\text{edge} + \frac{|V|}{|E|} \times n_\text{PE} \times (m_\text{memword}-m_\text{edge})\right)}
\]

The additional term can grow to be a problem if the access granularity is very large, if the average degree of the graph is low, or if the number of PEs in the system is very high. In the latter case there is however a limit: The edge list cannot be spread thinner than one edge per PE (if a vertex has zero neighbors on a PE, no memory request is issued). If the term $\frac{|V|}{|E|} \times n_\text{PE}$ reaches 1, the formula reduces to $T_\text{sys} \leq n_\text{FPGA} \times \frac{BW_\text{mem}}{m_\text{memword}}$: a whole memory word has to be read for each edge.

%Even more complex situations are possible. As an example, the HMC memory can be accessed in bursts of 1-8 words of size 128 bit, and the observed bandwidth is different depending on the size of access chosen.

\subsection{Network interface bandwidth constraint.}
\label{network_limit}
The PEs have to exchange messages among each other.

With the GraVF-M optimization, the amount of data transferred is reduced. Instead of one message per neighbor, one update is sent to each FPGA for all neighbors combined. On average, a vertex has $\frac{|E|}{|V|}$ neighbors. 
We assume that the network does not support a true broadcast operation, so each update has to be sent individually to all $n_\text{FPGA}-1$ remote FPGAs.  We also assume that the sending and receiving bandwidths are symmetrical, i.e. both equal to half the reported total interface bandwidth $\frac{BW_\text{if}}{2}$.

The average amount of data sent over the network per edge is $m_\text{update} \times (n_\text{FPGA}-1) \times \frac{|V|}{|E|}$.
$T_\text{FPGA}$ is the amount of edges processed per second on the sending FPGA. The limitation that the network interface of a given FPGA imposes on the performance is thus expressed by the inequality
\begin{equation*}
    T_\text{FPGA} \times m_\text{update} \times (n_\text{FPGA}-1) \times \frac{|V|}{|E|} \leq \frac{BW_\text{if}}{2}
\end{equation*}

Combining this with $T_\text{sys} = T_\text{FPGA} \times n_\text{FPGA}$ and solving for $T_\text{sys}$ results in:
\begin{equation} \label{eq:iflim}
\boxed{
T_\text{sys} \leq L_{if} = \frac{BW_\text{if}}{2 \times m_\text{update}} \times \frac{n_\text{FPGA}}{n_\text{FPGA}-1} \times \frac{|E|}{|V|}
}
\end{equation}

For GraVF, which sends unicast messages rather than broadcasting updates, the network interface limit has been calculated in \cite{engelhardt2018perfmodel} to be 
\begin{equation} \label{eq:iflim_old}
L_{if}(\text{GraVF}) = \frac{BW_\text{if}}{2 \times m_\text{message}} \times \frac{n_\text{FPGA}^2}{n_\text{FPGA}-1}
\end{equation}

Thus, the theoretical speedup of GraVF-M over GraVF is:

\begin{equation}
\boxed{
\text{Speedup} = \frac{L_{if}(\text{GraVF-M})}{L_{if}(\text{GraVF})} = \frac{|E|}{|V|} \times \frac{1}{n_\text{FPGA}} \times \frac{m_\text{update}}{m_\text{message}}
}
\end{equation}

Typically, $|E|$ is larger than $|V|$ by one to two orders of magnitude.
However, in for very large systems or extremely sparse graphs where the term $|E|/(|V| \times n_\text{FPGA})$ is smaller than 1, the optional filtering step described in section \ref{sec:filter} that sends updates only to FPGAs where neighbors reside ensures that the performance will not be reduced below that of GraVF. 
With our current implementation, $\frac{m_\text{update}}{m_\text{message}} = 1$ so this term does not affect performance.
% updates are actually about 30% smaller than messages, but the interface can only take one per cycle, so they both get rounded up to 128. Only if the two terms round to different multiples of 128bit is there any effect.

\subsection{Total network bandwidth constraint.}

The overall bandwidth of the network connecting the FPGAs is also limited. The cumulative data sent by all the FPGA boards cannot exceed it. As before, the average amount of data sent over the network per edge is $m_\text{update} \times (n_\text{FPGA}-1) \times \frac{|V|}{|E|}$. The total amount of edges processed per second in the whole system is $T_\text{sys}$.

The limitation introduced by the total network bandwidth can thus be expressed in the inequality

\begin{equation*}
T_\text{sys} \times m_\text{update} \times (n_\text{FPGA}-1) \times \frac{|V|}{|E|} \leq BW_\text{network}
\end{equation*}

Rearranging this equation give the network limit $L_{net}$:
\begin{equation} \label{eq:netlim}
\boxed{
T_\text{sys} \leq L_{net} = \frac{BW_\text{network}}{(n_\text{FPGA}-1) \times m_\text{update}} \times \frac{|E|}{|V|}
}
\end{equation}

Comparing this to the limits of GraVF found in \cite{engelhardt2018perfmodel}:

\begin{equation} \label{eq:netlim_old}
T_\text{sys} \leq L_{network} = \frac{BW_\text{network} \times n_\text{FPGA}}{ (n_\text{FPGA}-1) \times m_\text{message}}
\end{equation}

As the same reduction in amount of data crossing the network applies, we once again obtain the same speedup as for the network interface limit.

\begin{equation}
\label{eq:speedup}
\boxed{
\text{Speedup} = \frac{L_{if}(\text{GraVF-M})}{L_{if}(\text{GraVF})} = \frac{|E|}{|V|} \times \frac{1}{n_\text{FPGA}} \times \frac{m_\text{update}}{m_\text{message}}
}
\end{equation}

\paragraph{Other network topologies}

In these two network limits, only a simple network consisting of direct one-to-one communication between FPGAs is considered, as this corresponds to the platforms on which GraVF(-M) has been implemented so far. However, other topologies exist.

The reasoning on network and network interface constraints can be extended to hierarchical networks with more layers. This is commonly the case where e.g. multiple FPGAs are hosted on the same node and can communicate quickly with each other, but there is less bandwidth available to communicate with FPGAs on remote nodes. In this situation, the same equations apply, with the number of nodes in place of the number of FPGAs.

More generally, the techniques of network analysis can be applied by considering each FPGA as a source of $T_\text{FPGA} \times m_\text{update} \times \frac{|V|}{|E|}$ bits/s of data, which have to be broadcast to all $n_\text{FPGA}-1$ other FPGAs, and modelling whatever interconnect topology the system in question is equipped with.

\subsection{Determining the limiting factor}

Overall, taking into account all four factors, we can derive the predicted throughput of an FPGA:

\begin{equation} \label{eq:total}
T_\text{sys} = \min(L_{PE}, L_{mem}, L_{if}, L_{net})
\end{equation}

Since we would not want to computationally limit the system while there are resources available, in a first step we assume we use the maximum number of PEs available, fixing the value of $n_\text{PE/FPGA}$ in $L_{PE}$ and calling the resulting value $L_{PE_{max}}$.

$L_{PE_{max}}$ and $L_{mem}$ increase with $n_\text{FPGA}$, whereas $L_{if}$ and $L_{net}$ decrease with $n_\text{FPGA}$. 
Therefore, there are multiple candidates for the optimum value of $T_\text{sys}$. As the network limits are not defined for $n_\text{FPGA}=1$, it might be the best solution.
Otherwise, it will be obtained at the point where the network limits and one of the other limits intersect:

\begin{equation*}
    \min(L_{PE_{max}}, L_{mem}) = \max(L_{if}, L_{net})
\end{equation*}

This equality is solved for $n_\text{FPGA}$, and the resulting value of $T_\text{sys}$ compared to the single-FPGA performance to find out which is the real solution. Rounding to the nearest integer if necessary, we obtain the value of $n_\text{FPGA}$.

If memory or network is found to be the limiting factor, once the number of FPGAs is chosen, the number of PEs per FPGA can be lowered to save power. Substituting the values of $T_\text{sys}$ and $n_\text{FPGA}$ calculated in the previous step into equation \ref{eq:complim} gives a lower bound on the necessary number of PEs per FPGA to maintain the same performance:

\[n_\text{PE/FPGA} \geq \frac{T_\text{sys} \times CPE_{PE}}{n_\text{FPGA} \times f_\text{clk}}\]

In this manner, the framework chooses the optimal values for the number of FPGAs ($n_\text{FPGA}$) and the number of PEs per FPGA ($n_\text{PE/FPGA}$), optimizing first for throughput and then for power.

\section{Performance Evaluation}
\label{sec:results}

\subsection{Evaluation platform}

To evaluate the performance of the framework-generated designs, we use a system comprised of four Micron AC-510 modules on an EX-750 backplane. Each AC-510 module contains a Xilinx KU060 FPGA and a 4GB Hybrid Memory Cube connected with two half-width (x8) links with 15Gb/s signaling rate. The EX-750 backplane allows direct communication via PCIe x8 gen 3 link between the FPGAs without passing through the host.

To apply our analytical model described in section \ref{sec:model}, we need to determine the relevant parameters of the platform: the operating frequency $f_\text{clk}$, the PE throughput measure $CPE_\text{PE}$, the memory interface bandwidth $BW_\text{mem}$, the network interface bandwidth $BW_\text{if}$, and the total network bandwidth $BW_\text{network}$. 
Most of these factors are influenced by the implementation of the memory controller and PCIe endpoint provided by Micron, so we experimentally determine these values. 

The operating frequency $f_\text{clk}$ is determined by using the maximum supported by the memory controller interface, 187.5~MHz.

We derive the PE throughput measure $CPE_\text{PE}$ from the performance of a system with only one single PE using on-chip memory. As the network reduces to a single FIFO feeding the PE's output back into its input, no outside factors should impede performance in this situation. We measure $CPE_\text{PE} = 1.05$ for WCC, $CPE_\text{PE} = 1.10$ for BFS, and $CPE_\text{PE} = 1.42$ for PR.

For the memory interface bandwith, we use the GUPS benchmark included with the PicoFramework development environment. The peak experimentally determined available memory bandwidth is 21.7GiB/s, although for the configuration most closely matching our use case the observed bandwidth is only 8.1GiB/s (as memory bandwidth was never the limiting factor in our experiments, we did not invest further effort in optimizing this).

Determining the network parameters is more complicated. Communication over PCIe is wrapped by the PicoFramework into \emph{streams}, which are presented to the user like ordinary 128-bit wide first-word fall-through FIFOs with one end on either FPGA.
A separate stream has to be opened for communication between every pair of FPGAs in the system. The PCIe interface bandwidth $BW_\text{if}$ is divided among the streams, but the streams themselves also have an inherent maximum throughput of 128 bit every clock cycle. Either of these can be the limiting factor.

We implement a benchmark that continuously sends data to all other FPGAs in the system, and measure the cumulative sending and receiving bandwidth from all channels on one FPGA when sending data to 1, 2 or 3 other FPGAs. (The other FPGAs are also sending data to each other, in the same configuration as is used in GraVF/GraVF-M.)
To mirror our use case, the always-on sender and receiver are located in the 187.5MHz clock domain, and connected to the stream clock domain (200MHz) through an asynchronous FIFO. While one would think that this would result in less data transmitted than if the sender is directly in the stream clock domain, it does in fact lead to 20\% higher bandwidth transmissions in the 4-FPGA case, presumably due to congestion effects.
The results are shown in table \ref{tab:bw_bench}. We find that the network interface bandwidth is $BW_\text{if} = 11.7$GiB/s, but that performance is further restricted in the 2-FPGA and 3-FPGA case by the FIFO's interface bandwidth.

\renewcommand{\arraystretch}{1.3}
\begin{table}[h]
    \centering
    \begin{tabular}{|c|c|c|}
    \hline
        $n_\text{FPGA}$ & send (GiB/s) & receive (GiB/s) \\
        \hline\hline
        2          & 2.79 & 2.79 \\ 
        3          & 5.59 & 5.59 \\ 
        4          & 5.85 & 5.85 \\ 
        \hline
    \end{tabular}
    \caption{Observed network bandwidth}
    \label{tab:bw_bench}
\end{table}

As there is no slowdown at the maximum configuration, we are unable to determine $BW_\text{network}$ precisely, but we can determine that it is larger than the amount of data sent by all 4 FPGAs simultaneously, 23.4~GiB/s, and that it is never the limiting factor on this platform.

\subsection{Algorithms and datasets}

We implement kernels for three algorithms for use with our framework: Breadth-First Search, Weakly Connected Components, and PageRank. These algorithms are widely used by previous work and thus allow easy performance comparison. They also cover distant points on the spectrum of workloads, from the highly variable (BFS) to the regular (PR).

\paragraph{Breadth-First Search (BFS)}

BFS is a kernel that forms the base for a wide variety of applications, including the graph500 supercomputing ranking \cite{murphy2010introducing}. The most common variant, which we adopt, is one that visits all vertices in the graph in order to compute a minimum spanning tree, rather than a true search (which might be abandoned early upon finding a specific vertex) as the name suggests. Each vertex stores the ID of its parent vertex in the spanning tree.

In BFS, each vertex is active only once, during the superstep where it is first visited. Exactly one message is sent for each edge in the graph. The number of active vertices is highly variable over the course of execution, starting from a single active vertex in the first superstep, growing to encompass a large portion of the graph within a few hops, and finally reducing again to only a few remote vertices as it nears completion.

\paragraph{Weakly Connected Components (WCC)}

The WCC algorithm detects unconnected portions of a graph. Each connected set of vertices agrees on a common label by propagating the lowest seen vertex ID to all neighbors. (This algorithm is described in detail in section \ref{sec:progmodel}, where it is used as an example illustrating the kernels syntax accepted by GraVF-M.)

Many vertices will be active more than one superstep, especially those with high vertex IDs, as progressively lower vertex IDs are propagated. However, similarly to BFS the total amount of activity is limited by the diameter of the graph and wanes towards the end of the computation.

\paragraph{PageRank (PR)}

PageRank is a ranking algorithm that assigns a floating-point score to each vertex in the graph based on the score of its neighbors, computed over multiple iterations towards a fixed point. We have adapted the vertex-centric implementation from Pregel~\cite{malewicz2010pregel}.

During each superstep, each vertex contributes $\frac{1}{n}$th of its score from the previous iteration to each of its neighbors' scores, where $n$ is the number of (outgoing) neighbors that the vertex is connected to. Computation is run for 30 supersteps. All vertices are always active, and a message is sent over each edge of the graph in each superstep. PR is thus a heavier but more regular computation than the other two algorithms. The amount of computation per vertex per iteration  is also larger, consisting of multiple floating point operations which combine to give the pipelined kernel a latency of several hundred cycles.

\paragraph{Datasets}

We use two types of synthetic graphs to evaluate our framework.
Uniform graphs have edges distributed evenly, with most vertices close to the average degree. Edges are generated with equal probability for any pair of vertices.
Scale-free graphs have extremely skewed degree distributions following a power law. Most vertices have low degrees, but a few are highly connected. These datasets are obtained from the RMAT generator \cite{deepayan2004rmat}.

\paragraph{Measurement}

We evaluate performance using a common throughput measure for graph algorithms, \emph{traversed edges per second} (TEPS). It is not well-defined in the literature what counts as traversing an edge, especially for algorithms like WCC where vertices can receive new data from the same neighbor multiple times during the course of execution. Here, we consider that an edge is traversed every time information is transferred along it, i.e. when the scatter kernel sends a message. The overall system throughput $T_\text{sys}$ is calculated as the total number of messages sent over the course of the execution divided by the runtime.

\subsection{Experiments}

In these experiments, we aim to evaluate both the general performance of the framework-generated designs, and to compare the predictions of the performance model to the actual results.

\subsubsection{Multi-FPGA Scaling}

\begin{figure}[htbp]
\includegraphics[width=\linewidth]{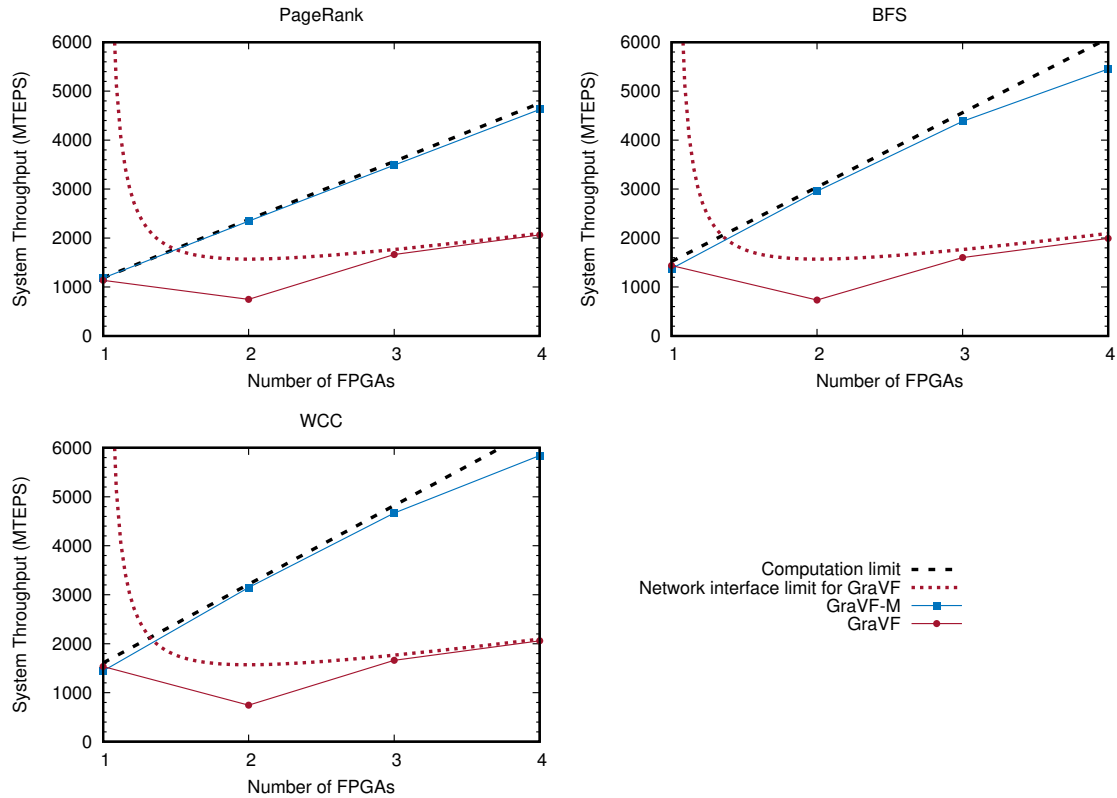}
\centering
\caption{Throughput for a multi-FPGA system\label{fig:pcie_througput}}
\end{figure}

To measure the effect of the improved architecture introduced in this paper, we compare the performance of GraVF-M to GraVF when scaling from one to four FPGA boards. Each FPGA is configured to use 9 PEs, with the use of off-chip HMC memory for edge data enabled. We use a uniform graph dataset, and partition it among the FPGAs using the greedy edge-based heuristic. Results are shown in Fig.~\ref{fig:pcie_througput}.

Overall, GraVF-M exhibits much better scaling performance. Although GraVF is still preferable when run on a single FPGA, GraVF-M achieves a $2.2-2.8 \times$ speedup when spreading computation across 4 FPGAs.

In the following sections we will examine some aspects of GraVF-M performance in more detail.

\subsubsection{Single-FPGA Performance}

\begin{figure}[h]
\centering
\includegraphics[width=\linewidth]{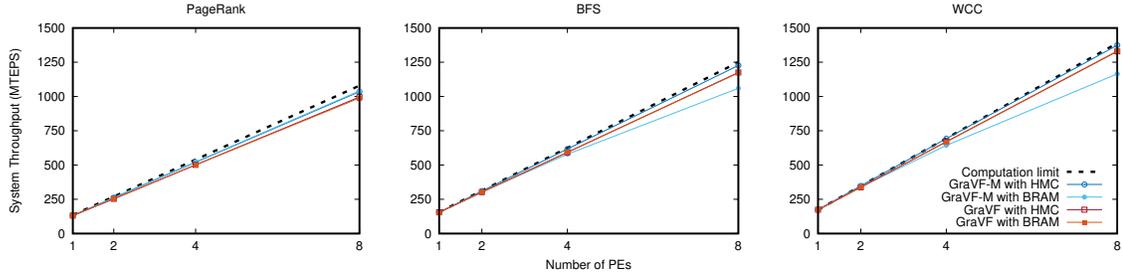}
\caption{GraVF-M vs. GraVF performance on uniform graph.\label{fig:scale_uniform}}
\end{figure}

We evaluate the PE performance when unconstrained by factors such as memory or inter-FPGA network. In this experiment, we only use a single FPGA, and we compare the \emph{GraVF} and \emph{GraVF-M} designs.
Fig. \ref{fig:scale_uniform} shows the performance of both designs for a uniform graph dataset.

In exchange for better multi-FPGA performance, the single-chip performance of GraVF-M is reduced compared to GraVF when using BRAM. The GraVF-M on-chip network is more prone to communication conflicts, as the broadcast mechanism results in the PEs sending nearly in lockstep. A single PE unable to accept more data also blocks all PEs from sending, whereas in GraVF PEs would still be able to send messages to other destinations, and only stop when chancing upon a message destined for the blocking PE. A large part of this performance can be regained when using the off-chip HMC memory, as the 64 in-flight requests per PE in effect provide extra buffer space.

\subsubsection{Effects of average degree}

The core GraVF-M architecture improvement introduces a dataset-dependent term $|E|/|V|$ when the performance is limited by the inter-FPGA network. Consequently, GraVF-M performance should be better for graphs with higher average degree.

\begin{figure}[h]
\centering
\includegraphics[width=0.5\linewidth]{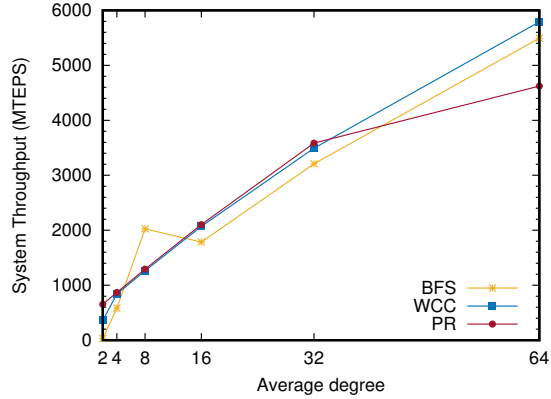}
\caption{Effect of average degree on GraVF-M performance.\label{fig:degree_effect}}
\end{figure}

We generate uniform graphs with a constant number of vertices and average degrees ranging from 2 to 64. Fig. \ref{fig:degree_effect} compares the performance of a 4-FPGA GraVF-M design using off-chip HMC memory with 9 PE per FPGA and with greedy-edge partitioning. The influence of the average degree ($|E|/|V|$) on performance can be clearly seen. At very low degree, the performance (especially of the BFS algorithm) is more variable as other effects dominate (BFS is sensitive to the shape of the graph).
At the highest degree shown (64), the computational limit of the PEs is reached before the network limit, especially for PR which has a higher CPE.

This reliance on the average degree does mean that GraVF-M does not benefit low-degree datasets such as road networks as much. On a subgraph of the Pennsylvania road network obtained from the SNAP database~\cite{snapnets}, which has an average degree of 2.8, we achieve a performance of 182 MTEPS for BFS, 720 MTEPS for PR and 663 MTEPS for WCC using 4 FPGAs.

\subsubsection{Latency effects}

\begin{figure}[h]
\centering
\includegraphics[width=0.6\linewidth]{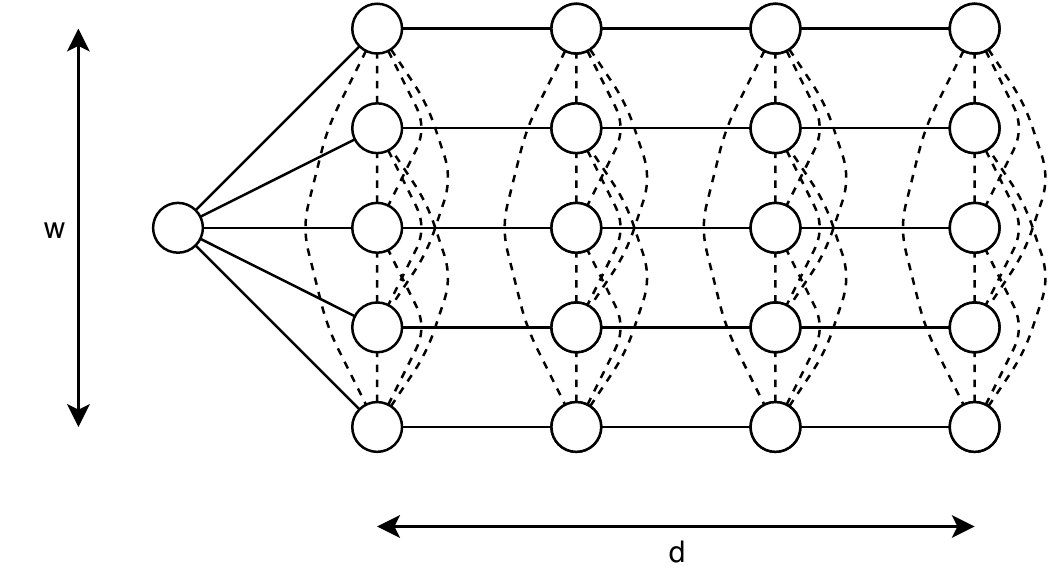}
\caption{Structure of the synthetic graphs used in this experiment. The single vertex on the left is selected as root for the BFS algorithm.\label{fig:fd_graph}}
\end{figure}

To estimate the efficiency of the barrier synchronization, we create a set of synthetic graphs with the structure depicted in Fig.~\ref{fig:fd_graph}, which allows us to precisely control the number of supersteps ($d$ + 1) and the number of vertices active in each superstep ($w$) for the BFS algorithm (the most latency-sensitive of our algorithms). The solid edges represent the BFS minimum spanning tree, and the additional edges (drawn dashed) between vertices of same rank serve to achieve a higher average degree without affecting when a vertex will be active.

In a first step, we run BFS on a line graph with 16385 vertices (i.e. the graph obtained for $w=1, d=16384$ and with no additional edges). This graph will have only a single vertex active each superstep, for 16385 consecutive supersteps. As the amount of work per superstep is negligible, the time used per superstep is entirely attributable to synchronization latency. From this experiment we obtain a latency of 676 cycles/superstep, or 3.6µs, for a system of 4 FPGAs with 9 PEs each, using HMC (i.e. the same system as in Fig.~\ref{fig:pcie_througput}). This can be compared to the round-trip time between two FPGAs in this system, which is 1.6µs.

\begin{figure}[h]
\centering
\includegraphics[width=0.65\linewidth]{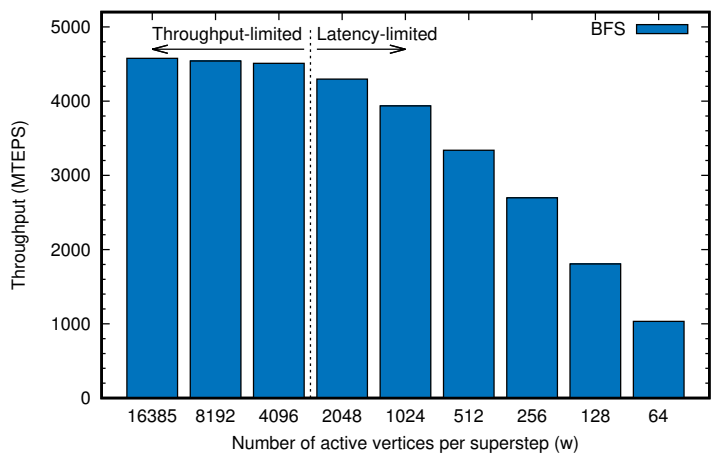}
\caption{Throughput obtained for synthetic graphs as in Fig.~\ref{fig:fd_graph} when increasing $d$ and decreasing $w$ while keeping the total vertex and edge count constant.\label{fig:fd_result}}
\end{figure}

In a second step, we create a series of these synthetic graphs with a constant number of vertices, but distributed across an increasing number of supersteps. The average degree for the graphs is set to 64, and the number of vertices active in each superstep $w$ is decreased from 16384 to 64, while $d$ is increased from 1 to 256 supersteps.
Fig.~\ref{fig:fd_result} shows the throughput on this series of graphs, for the same system of 4 FPGAs with 9 PEs each, using HMC. It can be seen that latency becomes a significant factor starting at 2k active vertices per superstep, which corresponds to 128k edges traversed per superstep. As the number of supersteps increases and the amount of work per superstep decreases, the time per superstep converges towards the synchronization latency previously determined.

\subsubsection{Partitioning}

In this experiment we examine the effectiveness of the partitioning methods. Partitioning is more challenging for scale-free graphs, so we use the RMAT-generated graphs. The system is configured to use HMC memory.

\begin{figure}[h]
\centering
\includegraphics[width=\linewidth]{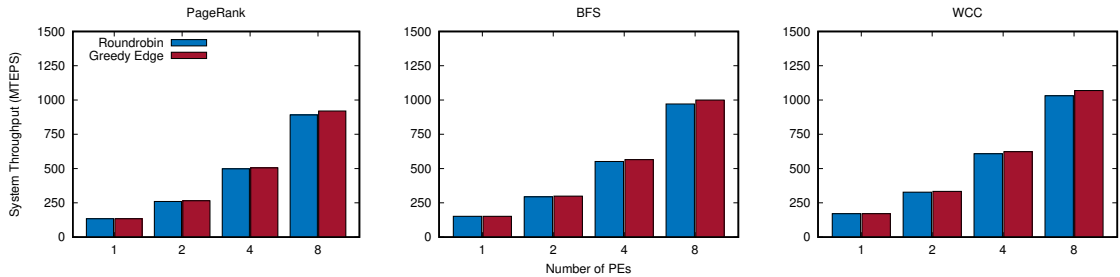}
\caption{Round-robin vs. greedy edge-based partitioning.\label{fig:partition_1fpga}}
\end{figure}

Fig. \ref{fig:partition_1fpga} shows the comparison of the round-robin and greedy partitioning strategies on single-FPGA systems of 1 to 8 PEs. The greedy strategy shows a slight advantage, but the round-robin partitioning is not significantly poorer-performing. Greedy edge-based partitioning was adopted as the default strategy.

\begin{figure}[h]
\centering
\includegraphics[width=\linewidth]{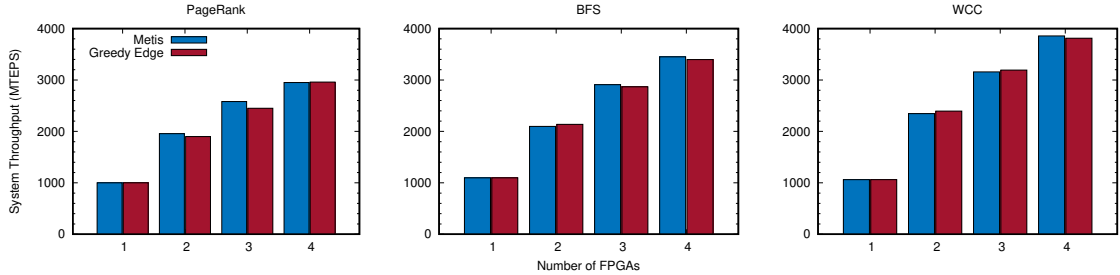}
\caption{Greedy edge-based partitioning vs. METIS.\label{fig:partition_4fpga}}
\end{figure}

Fig. \ref{fig:partition_4fpga} compares the performance when scaling across multiple FPGAs using location-agnostic partitioning from the greedy edge-based heuristic, or a high-quality partition obtained from the METIS tool \cite{karypis1998metis} that attempts to minimize the number of edges crossing FPGA boundaries. As METIS runtime is several thousand times higher than the runtime of the algorithms under investigation, it should be viewed more as an upper bound on the effects of partitioning quality rather than as a realistic option. The greedy edge-based heuristic is able to achieve results within 5\% of METIS.

\subsection{Comparison to other frameworks}

As is noted by \cite{oguntebi2016graphops}, the commonly-used throughput measure MTEPS (Millions of Traversed Edges Per Second) is not always comparable across different works, because the notion of what constitutes `traversing' an edge is not well-defined. The related works examined in this section were not described in enough detail to ascertain the authors' definition of traversal; as such we can merely take the reported numbers at face value and assume that it matches our definition.

\subsubsection{Multi-FPGA frameworks}

ForeGraph \cite{dai2017multifpga} is the only multi-FPGA framework to compare our work with. Their work is evaluated by simulating a 4-FPGA platform similar to ours (memory bandwidth of 19.2~GB/s vs. 21.7GiB/s, network bandwidth 1.4~GiB/s vs. 1.5~GiB/s), and with the same benchmark algorithms, on a social graph of comparable edgefactor (35, vs. 32 for the dataset that our results were obtained on).
% edgefactor 70 (TW)
Table \ref{tab:gravf_vs_foregraph} shows the peak performance reported for ForeGraph and that obtained for our work. GraVF-M achieves performance $2.5-3.8\times$ that of the closest comparable work.

\begin{table}[h]
% \noindent\resizebox{\columnwidth}{!}{
\begin{tabular}{c|c|c}
\hline
Algorithm & ForeGraph & GraVF-M \\
\hline
PageRank & 1856 & 4623 \\
BFS & 1458 & 5493 \\
WCC & 1727 & 5791 \\
\hline
% \vspace{0.3em}
\end{tabular}
\caption{Peak performance obtained for different algorithms (MTEPS) \label{tab:gravf_vs_foregraph}}
% }
\end{table}

\subsubsection{Single-FPGA frameworks}

We can also compare the performance of single-FPGA frameworks to the per-FPGA performance of GraVF-M by dividing the throughput by the number of FPGAs. The following works report throughput numbers on one of our benchmark algorithms in a compatible format:

\cite{zhouprasanna2016edge} achieves 747-1068~MTEPS for the WCC algorithm, on a platform with a memory bandwidth of 19.2~GB/s and on datasets with edgefactor 2-14. Our work has a throughput of 1447~MTEPS per FPGA.

\cite{zhouprasanna2017vertexedge} reports executing BFS at 670 MTEPS, on a hybrid platform comprised of a CPU with memory bandwidth 30~GB/s and an FPGA with memory bandwidth 12.8~GB/s, on synthetic graphs with edgefactors of 4 to 16, compared to 1373~MTEPS per FPGA for GraVF-M.

\cite{zhouprasanna2018edgecentric} includes an evaluation using PR, on a platform with 60~GB/s memory bandwidth with 1116-2487~MTEPS on social graphs with edgefactors 2-35. Our work has a per-FPGA throughput of 1155~MTEPS on this algorithm.

\cite{oguntebi2016graphops} shows throughput of approximately 110~MTEPS for PR, but it is very likely that their definition of edge traversal differs. Their SpMV algorithm achieves 750~MTEPS and PageRank can be implemented as a succession of sparse matrix-vector multiplications, so the performance of the two should not differ this drastically under our definition.

Other works' performance is even more difficult to compare, as they use different algorithms for their evaluation or report their numbers in speedup against some software baseline (or in the case of \cite{zhou2017tunao}, GPU).

\section{Conclusion}
\label{sec:conclusion}

In this work we have presented the GraVF-M framework, which generates graph processing designs optimized for multi-FPGA platforms. This framework greatly simplifies the implementation of FPGA based accelerators for vertex-centric graph processing algorithms, while offering attractive performance. Our proposed architecture allows the familiar message-passing paradigm to be maintained, while exploiting structural features of the programming model to transfer portions of the computation to the receiver in order to minimize inter-FPGA network traffic.

A 4-FPGA system generated by GraVF-M reaches 4.6-5.8 GTEPS, which compares favorably to other works in the literature. It is also up to 94\% of the theoretical peak performance projected by our performance analysis. The performance model we developed can be used to evaluate the likely performance of platforms based on their memory and network interface capabilities before proceeding to implementation.

\begin{acks}
This work was supported in part by the \grantsponsor{GRF17245716}{Research Grants Council of Hong Kong}{} (Project GRF \grantnum{GRF17245716}{17245716}), and the \grantsponsor{croucher}{Croucher Foundation}{} (\grantnum{croucher}{Croucher Innovation Award 2013}).
The authors would also like to thank Micron for the loan of the Pico SC-6 Mini evaluation platform. 
\end{acks}

\bibliographystyle{ACM-Reference-Format}
\bibliography{master.bib}

\end{document}